\documentclass[twocolumn]{aastex701}

\DeclareRobustCommand{\VAN}[3]{#2}
\let\VANthebibliography\thebibliography
\def\thebibliography{\DeclareRobustCommand{\VAN}[3]{##3}\VANthebibliography}
\usepackage[percent]{overpic}
\usepackage{xcolor} 
\usepackage{placeins}
\usepackage{soul}
\newcommand\N[1]{{\color{blue} \bf #1}}
\newcommand\M[1]{{\color{green} \bf #1}}

\begin{document}

\title{The Age of the R127 \& R128 Clusters: Implications for the LBV}

\correspondingauthor{Mojgan Aghakhanloo}
\email[show]{mvy4at@virginia.edu}

\author[0000-0001-8341-3940]{Mojgan Aghakhanloo}
\affiliation{Department of Astronomy\\
 University of Virginia\\
Charlottesville, 22904, VA, USA}
\email{}

\author[0000-0003-1599-5706]{Jeremiah W.~Murphy }
\affiliation{Department of Physics\\
Florida State University\\
77 Chieftan Way\\
Tallahassee, 32306, FL, USA}
\email{}

\author[0000-0001-5510-2424]{Nathan Smith}
\affiliation{Steward Observatory\\
 University of Arizona\\
 933 N. Cherry Ave.\\
 Tucson, 85721, AZ, USA}
\email{}

\author[0000-0001-8878-4994]{Joseph Guzman}
\affiliation{Department of Physics\\
Florida State University\\
77 Chieftan Way\\
Tallahassee, 32306, FL, USA}
\email{}




\begin{abstract}

We infer the age of the R127 and R128 clusters in the Large Magellanic Cloud (LMC) using Strömgren photometry from the literature and the age-dating algorithm, \textit{Stellar Ages}.  Analysis using single-star evolutionary models shows a substantial discrepancy between the relative numbers of bright blue stars and lower-mass stars as compared to expectations from a Salpeter mass function, and yields a younger age for the brightest blue stars than for the rest of the cluster. This inconsistency reflects an emerging trend among young clusters in the Local Group.  In general, the resolution may be binary evolution or very rapid rotation, although in the specific case of the R127 and R128 clusters, unknown incompleteness in the data may also affect the relative numbers of low- and high-mass stars. The discrepancy grows toward fainter magnitudes, suggesting that the dataset is likely incomplete. However, when the five brightest stars are excluded, the observed and expected counts become consistent, demonstrating that the brightest stars are peculiar. These findings have direct implications for the luminous blue variable (LBV) R127, which is the only confirmed LBV in the LMC located within a young stellar cluster. LBVs have traditionally been considered products of single-star evolution, although there is growing recognition that binary interactions may play a critical role in their evolution. A more complete dataset, particularly deeper imaging with the Hubble Space Telescope, is needed to confirm whether the apparent absence of coeval stars arises solely from observational incompleteness or the broader trend of inconsistency in young cluster modeling.


\end{abstract}



\section{Introduction} \label{sec:intro}
Luminous blue variables (LBVs) are a short-lived and enigmatic evolutionary phase of massive stars after they leave the main sequence. They are characterized by extreme episodic mass loss, variability in brightness, and outbursts that can eject significant portions of their outer layers \citep{H94,S26}. These massive stars serve as valuable laboratories for understanding the physics of stellar winds, mass loss, and instabilities in high-mass stars \citep{S06,S14}. However, despite their importance in the massive star life cycle, many aspects of LBV evolution remain poorly understood. Key uncertainties include the mechanisms driving their extreme behavior, total mass loss, exact evolutionary pathways, and connection to the formation of both Wolf–Rayet (WR) stars and core-collapse supernovae \citep[SNe;][]{S11,S14}.

The single-star view has been that stars above 30--40 M$_{\odot}$ go through an LBV phase during the transition from core hydrogen burning to helium burning \citep{C98}. In this brief phase, they experience eruptive mass loss to transition from a hydrogen-rich star to a hydrogen-poor WR star. However, recent observations pose challenges to the single-star view of LBVs: (1) Some Type IIn SNe have direct LBV progenitors \citep{GL09,M13,S10,S22,B22,J22}. This challenges the traditional view of LBVs as single massive stars transitioning to core helium burning.  (2) LBVs, due to their expected short lifetimes, should remain close to their birth sites in young stellar clusters containing comparably massive O-type stars.  Yet, \cite{S15} found that most LBVs in the Large Magellanic Cloud (LMC) are isolated from O-type stars. This challenges their single-star evolutionary origins and suggests a binary scenario as a more likely explanation.

The discrepancies between observations and the single-star evolutionary scenario suggest that binary evolution may play a dominant role in the LBV phase of massive stars. This aligns with the growing recognition that binary interactions are a critical factor in the evolution of massive stars. Studies by \cite{S12} and \cite{M17} show that approximately 70\% -- 100\% of massive stars have a companion close enough to undergo mass transfer at some point in their evolution. These interactions can significantly affect both stars through mechanisms such as mass transfer, common envelope evolution, and tidal forces. As a result, the single-star evolution model may not apply when a close companion is present. Certain evolutionary stages, such as LBV and WR-star stages, may be primarily achieved through binary evolution with mass exchange, rather than through single-star evolution \citep{V07,D17}.

If LBVs are products of binary evolution, they could be either mass gainers, where the primary mass donor has already exploded as a supernova (SN), imparting the companion with significant net motion, or rejuvenated either by mass accretion (without a significant kick from a companion’s SN) or by a stellar merger \citep{J14,S15,S18}. These evolutionary pathways may explain why LBVs appear isolated from O-type stars. \cite{A17} investigated this hypothesis by developing models for single-star and binary scenarios in the context of cluster dissolution.
While their analytical single-star model successfully explains the observed spatial separation of O-type stars in single-star scenarios, it cannot account for the more extreme isolation observed in LBVs. They found that the large separations of LMC LBVs from other O-type stars require either significantly higher dispersal velocities, such as those associated with runaway stars, or longer lifetimes, as might be expected for stars formed through binary interactions, either as mass gainers or as products of stellar mergers. 

There are 15 LBVs \citep{S15} in the LMC. At least 33\% of LMC LBVs exhibit radial velocities exceeding 25 km s$^{-1}$ relative to their local environment, suggesting that they are runaway/walkaway stars \citep{A22}. Indeed, despite their high luminosity, the overall spatial distribution of LBVs resembles a larger population of less massive, aging B-type supergiants \citep{S19}. Among the LMC LBVs, R127 \citep[HDE 269858;][]{S83} stands out as the only one known to reside within a well-defined stellar cluster (R143 is in the outskirts of 30 Doradus \citep{P93} but not in the R136 cluster, while S Dor and R85 are in an older loose association spread across $\sim$200 pc \citep{M00}). As such, we expect R127 to have a low relative space velocity.  Indeed, its measured radial velocity is 1.4 km s$^{-1}$ relative to its local environment.  \citet{M09} also estimated the radial velocity of R127 using lines with P Cygni features, resulting in slightly higher values. This remains consistent with \citet{A22}, accounting for the influence of the P Cygni features on the measurements.  Furthermore, \citet{D24} determined a transverse peculiar velocity for R127 of 6.1 $\pm$ 6.9 km s$^{-1}$.  These observations reinforce the conclusion that R127 has a low total velocity, which in turn naturally explains why R127 is one of the few LBVs to be located within a young open cluster.

The R127 cluster is a small Trapezium-like group consisting of at least 14 members with masses ranging from $\sim$ 15 to 85 M$_{\odot}$. LBV R127 (hereafter R127), the brightest cluster member, has a very high luminosity \citep[$>$10$^{6.1}$ $L_{\odot}$;][]{S83,W89,L98,V01}, which according to single-star evolutionary tracks corresponds to an initial mass of 80--90 M$_{\odot}$. R127 is therefore among the most luminous LBVs known, similar to the Galactic case of AG Car, and is an original defining member of the classical high-luminosity LBVs.  With such a high initial mass appropriate for a single star, it is expected to have a short lifetime of only 3 Myr. 

The R127 cluster is next to the neighboring R128 cluster. The two clusters are resolved as compact stellar groups with a projected separation of $\sim$ 20 arcseconds. 
The R128 cluster has at least 33 members, including the supergiant R128 \citep[hereafter R128; HDE 269859;][]{H03}. \citet{V98} reported photometric fluctuations in R128, with a total range of approximately 0.3 mag over timescales of a week to a month between 1983 and 1990. However, \citet{V01} suggested that R128 is unlikely to be an S Dor variable and instead classified it as a normal $\alpha$ Cyg variable. 

The primary goal of this paper is to determine the age of the stellar population in the R127 \& R128 clusters and, by extension, constrain the age of R127.
Establishing this age provides crucial insight into the evolutionary stage of the LBV and whether its current properties can be reconciled with single-star models of massive star evolution. R127 is particularly important in this regard, as it is the only LMC LBV that resides in its birth cluster. Thus, if any LMC LBV could plausibly represent the single-star evolutionary pathway, R127 is the most likely candidate. We assess whether the population-level age constraints inferred from the environment are consistent with single-star expectations, thereby testing whether R127 is more likely the product of single-star evolution or binary interaction, such as a mass gainer or merger.
Section~\ref{sec:obs} describes the photometric data utilized in this study, while Section~\ref{sec:method} outlines the age-dating algorithm and methodology. The results are presented and discussed in Section~\ref{sec:result}. In Section~\ref{sec:summary}, we conclude with a summary. 

\section{Observations}\label{sec:obs}

\cite{H03} observed a 1-arcminute-radius region around the R127 \& R128 clusters using the ESO New Technology Telescope (NTT) with the SUSI imager and the Strömgren \( u, v, b, \) and \( y \) filters. 
They calibrated the \( y \) magnitudes using literature values and inferred the \( b \) magnitudes from 
\( b - y \) colors. Since they did not obtain complete standard star observations across all filters and air masses, the color calibration may contain systematic offsets.

\begin{figure}
\includegraphics[width=\linewidth, trim=0 0 0 0, clip]{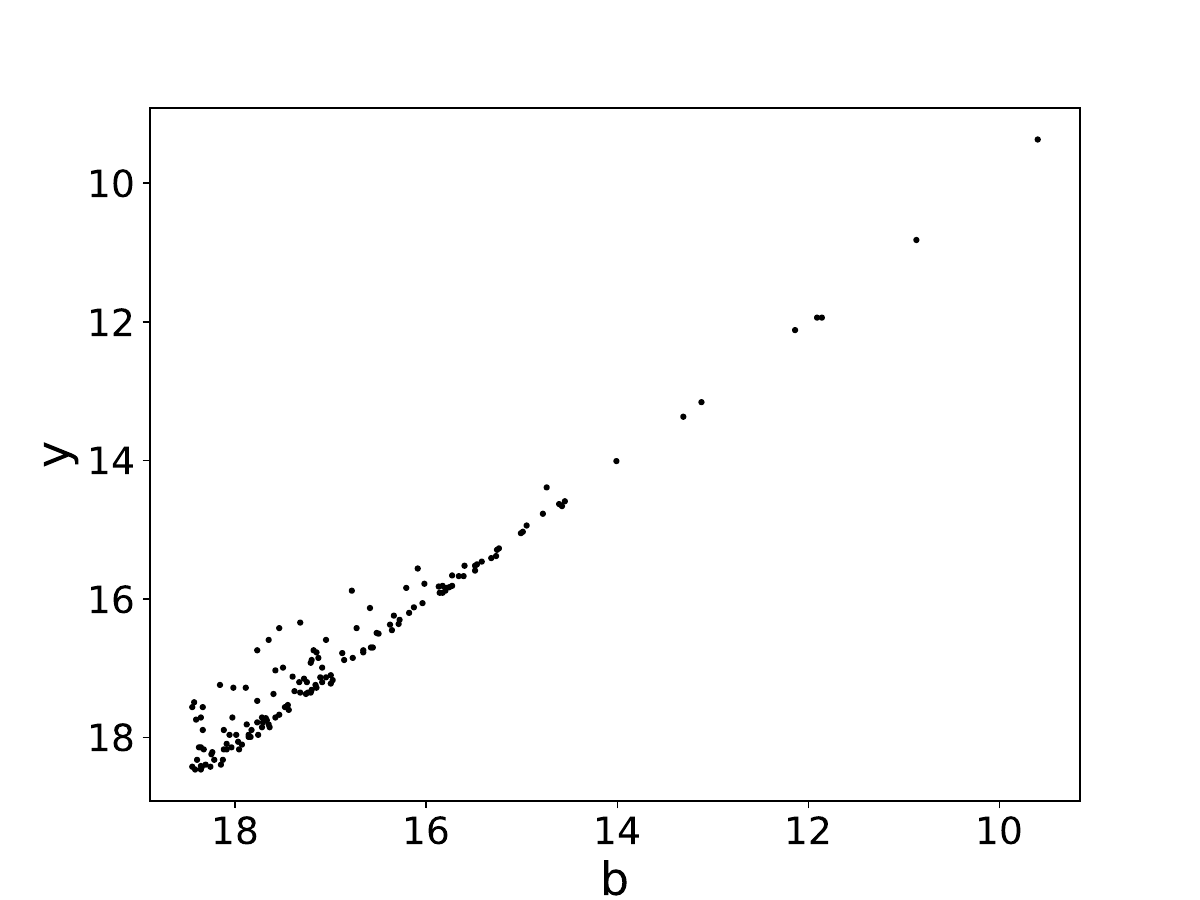}
\caption{Strömgren photometry for 147 stars associated with the R127 \& R128 clusters and their surrounding field stars. The data are obtained from Figure 8 of \citet{H03} and represent the most complete dataset available for this region. R127 is the brightest star in the region.
\label{fig:MalayeriData}}
\end{figure}

\cite{H03} resolved the tightly packed R127 \& R128 clusters using the MCS deconvolution algorithm, identifying 14 and 33 stellar components, respectively. They selected stars as cluster members based on their spatial association with the compact cores of R127 \& R128 clusters. They excluded some sources, particularly very faint stars (which, based on their tables, are likely fainter than $\sim$18 mag) or those blended with brighter neighbors, from the final photometric catalog due to low signal-to-noise or deconvolution residuals. 
While the complete dataset is not available online, we retrieved the data using available online sources. 
From their Figure 8, we extracted photometry for 204 stars. In contrast, their Tables~1--3 list only 78 stars, 15 of which lack measurements in \( y \) or \( b \). Thus, Figure~8 alone provides 141 additional data points. However, \citet{H03} reported obtaining photometry of 291 sources in total, leaving photometry for 87 stars unavailable, most likely due to the absence of both \( y \) and \( b \) magnitudes.
Because the full dataset is missing, we can only associate photometry with positions for stars listed in the tables, which prevents us from analyzing the R127 \& R128 clusters separately. This limitation does not affect our conclusions, however, as no kinematic study has demonstrated whether the two are distinct clusters or remnants of a single cluster that became partially divided through internal dynamical evolution or external tidal effects. 
Fig.~\ref{fig:MalayeriData} presents the \( y \)- versus \( b \)-band photometry for 147 observed stars with magnitudes ranging from approximately 9 to 18.5. For this study, we only utilize the data in the \( y \) and \( b \) filters, and faint sources ($>18.5$) are excluded as they are not used in the age-dating analysis; see Section~\ref{sec:method} for more details.

\begin{figure}
\includegraphics[width=\linewidth, trim=0 0 0 0, clip]{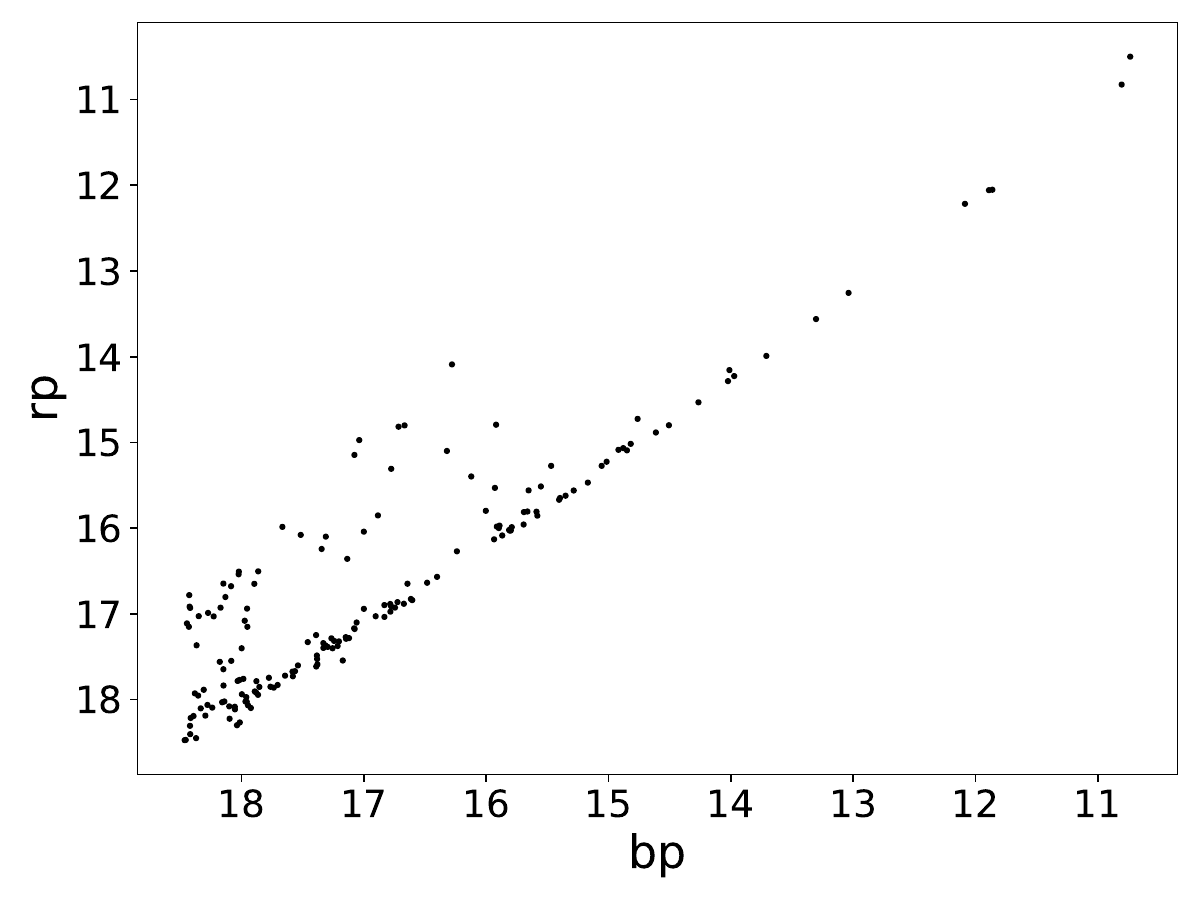}
\caption{Gaia DR3 data within 1\arcmin of R127. 349 out of 887 stars lack magnitude information. The plot shows only stars brighter than magnitude 18.5, comprising 169 sources. A gap around the 16th magnitude suggests that the Gaia DR3 data are incomplete along the main sequence in this region and therefore unsuitable for age dating the environment of R127. 
\label{fig:GaiaData}}
\end{figure}

The same region has also been observed in Gaia  Data Release 3 \citep[DR3;][]{G23}. Note that the Gaia  query covers a circular 1\arcmin region centered on the LBV, while the NTT field is a square area not centered on the LBV. Fig.~\ref{fig:GaiaData} displays the \texttt{phot-rp-mean-mag} versus \texttt{phot-bp-mean-mag} plot for stars brighter than 18.5 magnitude within a 1-arcminute radius around R127. A total of 887 stars were detected in this region, but only 538 of them have both \texttt{phot-rp-mean-mag} and \texttt{phot-bp-mean-mag} measurements, and just 169 are brighter than 18.5. Even though the Gaia DR3 data include more stars overall in this region, the NTT data are more complete in terms of cluster members, as they contain more stars along the main sequence. Notably, stars around magnitude 16 are missing photometric data, likely due to factors such as crowding, binarity, etc. This incompleteness highlights the limitations of the Gaia DR3 dataset for this analysis. 
Therefore, the NTT observations provide a more complete dataset, making them more suitable for the detailed analysis performed in this study. A detailed explanation of the algorithm used for age dating and its reliance on a complete dataset is provided in the next section.

\section{Age-dating algorithm (\textit{Stellar Ages})}\label{sec:method} 

To assess the age distribution of the R127 \& R128 stellar clusters, we apply the \textit{Stellar Ages} algorithm, a Bayesian statistical framework developed to infer the star formation history of resolved stellar populations using multiband photometry. \textit{Stellar Ages} provides a posterior probability distribution over physical parameters including age ($t$), metallicity ([M/H]), and rotation ($v_{ini}$), by modeling the observed magnitudes of stars in the context of an underlying stellar population.  In order to infer ages, this algorithm adopts predictions of stellar evolutionary models, as discussed in more detail below in Section~\ref{sec:model} .

At its core, \textit{Stellar Ages} answers the statistical question: Given the magnitudes of all stars in a population, what is the most likely combination of age, metallicity, and rotation that explains the data? Each star contributes to the overall likelihood by comparing its observed photometry with predictions from stellar evolution models for a given set of physical parameters. Crucially, this inference is not done in isolation. Each star’s likelihood is computed in the context of a mixture model that represents a weighted sum of many potential stellar populations with different ages, metallicities, and rotation rates.

This mixture model accounts for the possibility that the observed population is not homogeneous, and allows the algorithm to infer population parameters over a discrete 3D grid. To efficiently explore this space, \textit{Stellar Ages} employs a Gibbs sampling algorithm that iteratively updates both the latent population assignments of individual stars, and the weights $w_{t,z,v}$ that characterize the star formation history.

Each point on the model grid (indexed by $t$, $z$, and $v$) defines a subpopulation with known photometric predictions. For every star, the algorithm evaluates the likelihood of its observed magnitudes under each subpopulation. A latent label $r_{i} = (t',z',v')$ is assigned to each star, denoting its most likely source population. These labels are updated during Gibbs sampling, along with the weights $w_{t,z,v}$, from a Dirichlet prior. The resulting weights represent the fractional contributions of each model subpopulation to the total observed population and form the key output: an inferred distribution of ages, metallicities and rotations across the observed field.

Because \textit{Stellar Ages} is a population-level aware inference method, a consistent stellar catalog is critical. The algorithm does not infer properties of individual stars in isolation. Rather, it asks: What properties of this star, within the context of the entire population, make this individual star’s observed photometry most probable?

This implies that the algorithm is sensitive to the overall shape of the observed magnitude-magnitude diagram. If the magnitude-magnitude diagram were incomplete, e.g. missing faint stars due to incompleteness or crowding, then the inferred weights (and thus the inferred ages) may be skewed. Incomplete data can lead the algorithm to over- or underrepresent certain evolutionary phases, biasing the inferred properties of stars across the magnitude-magnitude diagram. However, the method does not require a complete catalog to yield meaningful results. In cases where only a handful of evolved stars are detected, these objects alone can strongly constrain the likely age of their parent population, albeit with correspondingly reduced sensitivity to other phases of the magnitude-magnitude diagram.

For this reason, a star's inferred age does not depend solely on its own $y$ and $b$ magnitudes but also depends on the distribution of lower-mass stars. For example, a sparsely populated main sequence can make a young age less likely, since a young population would predict numerous lower-mass companions that are not observed. Therefore, photometric incompleteness can systematically alter the inferred population age and should be considered under this method. In practice, the robustness of \textit{Stellar Ages} lies in its ability to integrate whatever evolutionary information is available into a consistent probabilistic inference. For further methodological details and validation, see \citet{guzman2025a}.

\subsection{Evolutionary models}\label{sec:model} 
Stellar evolutionary models, such as MESA Isochrones and Stellar Tracks \citep[MIST;][]{C16, D16} and PARSEC \citep{B12}, provide valuable frameworks for studying stellar evolution. These models were selected for their coverage of stellar evolution phases and widespread use in the field. However, they are primarily developed for single stars and do not account for binary interactions, which are common among massive stars and could significantly influence their evolution. As such, caution is warranted when interpreting the outputs of the algorithm, with careful attention to inconsistencies and discrepancies. In particular, the physics behind the LBV phase remains poorly understood, and therefore none of the currently available models explicitly include this critical evolutionary stage. The inferred age for any individual LBV is therefore highly suspect. Since \textit{Stellar Ages} estimates ages based on the overall distribution of stars, however, the absence of an LBV phase in these models should not strongly influence the age we infer for the overall stellar population.

\begin{figure*}[t]
    \centering
    \begin{overpic}[width=0.328\textwidth, trim=10 0 10 10, clip]{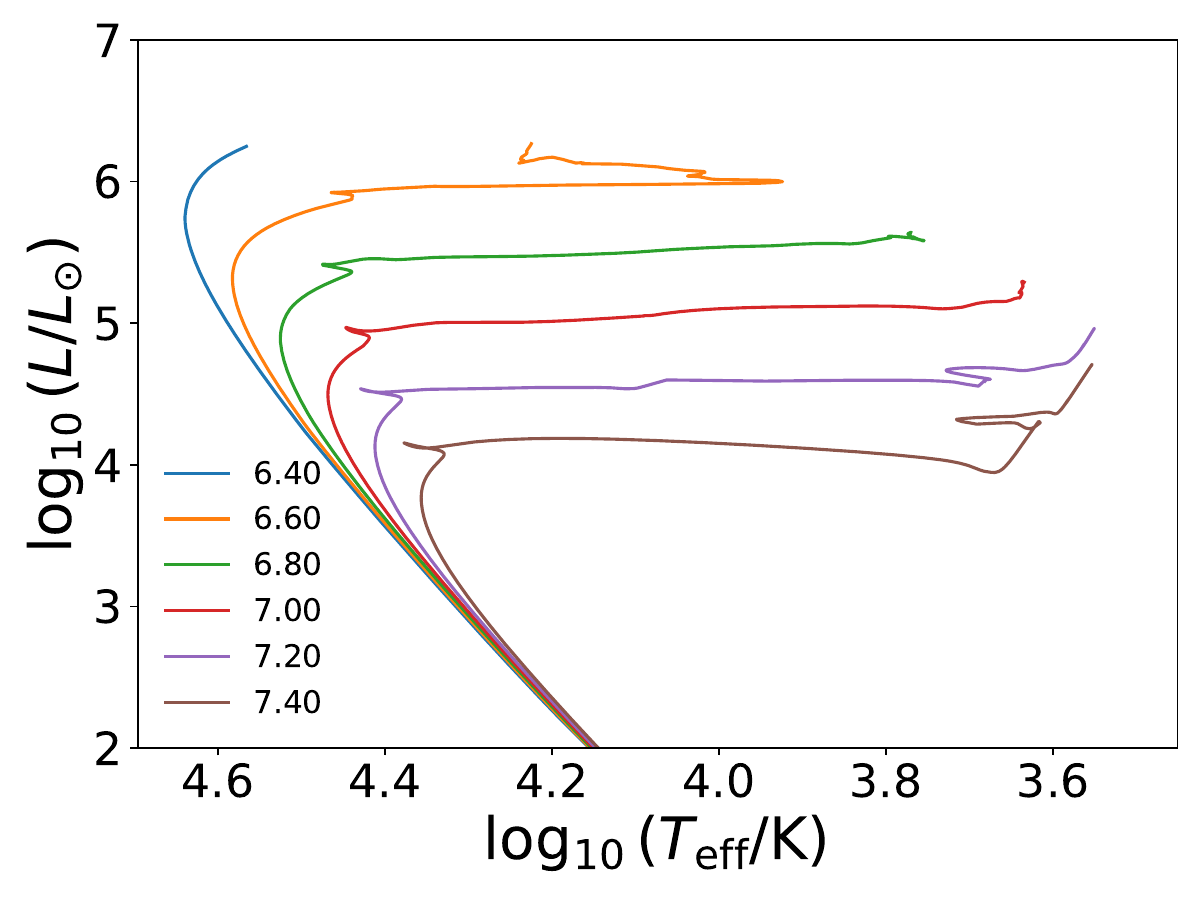}
        \put(85,73){\makebox(0,0)[lt]{\textcolor{black}{\footnotesize MIST}}}
    \end{overpic}
    \hfill
    \begin{overpic}[width=0.328\textwidth, trim=10 0 10 10, clip]{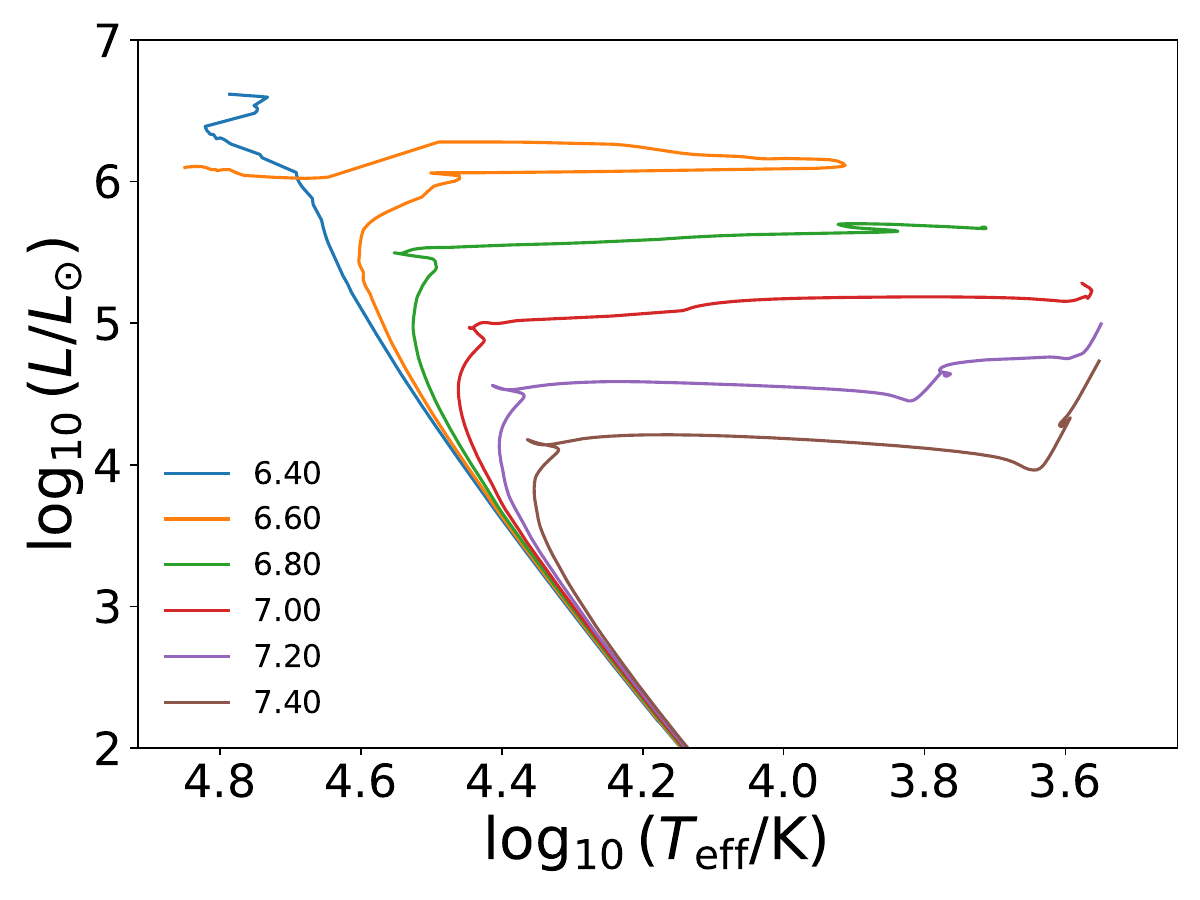}
        \put(38,73){\makebox(0,0)[lt]{\textcolor{black}{\footnotesize MIST $\Omega_{ZAMS}/\Omega_{crit}=0.4$}}}
    \end{overpic}
    \hfill
    \begin{overpic}[width=0.328\textwidth, trim=10 0 10 10, clip]{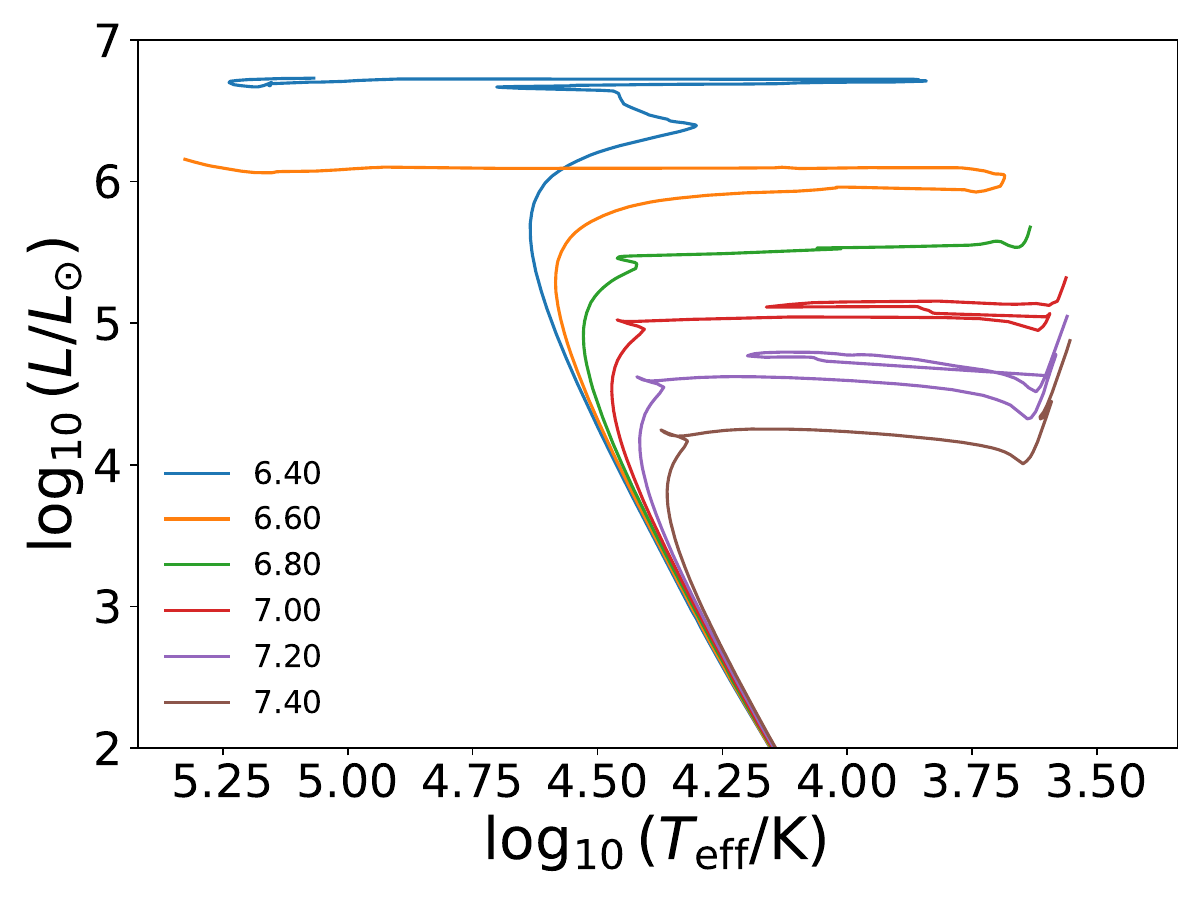}
        \put(78,73){\makebox(0,0)[lt]{\textcolor{black}{\footnotesize PARSEC}}}
    \end{overpic}
    \caption{Single-star evolutionary isochrones for LMC-like metallicity ([M/H] = $-0.40$) at different ages. The left panel represents MIST models, the middle panel shows MIST models with a $\Omega_{ZAMS}/\Omega_{crit}=0.4$, and the right panel shows PARSEC isochrones without rotation. The isochrones are very similar during the main-sequence phase, which is why we obtain consistent results when using different sets of isochrones (see the Appendix~\ref{sec:appendix}).}
    \label{fig:models}
\end{figure*}

Fig.~\ref{fig:models} presents single-star isochrones for LMC-like metallicity ([M/H] = $-0.40$) across a range of ages. We use MIST v1.2 models, which adopt a solar metallicity of $Z_\odot = 0.0142$. While our analysis incorporates a range of metallicities between [M/H] =$-0.4$ and 0.2, here we show isochrones for [M/H] =  $-0.4$ as a representative example. The left panel shows MIST isochrones for ages corresponding to logarithms between 6.40 and 7.40. The middle panel displays models incorporating solid-body rotation initialized at the zero-age main sequence (ZAMS), with a ratio of $\Omega_{ZAMS}/\Omega_{crit} = 0.4$. The right panel illustrates PARSEC v1.2S isochrones without rotation. The stellar age-dating algorithm used in this study is compatible with any of these isochrones.

Despite the challenges mentioned above, MIST and PARSEC remain among the most widely used and comprehensive evolutionary isochrones currently available for massive stars.  In particular, accounting for rotation in MIST models produces only a minor effect on the overall stellar distribution in the magnitude-magnitude diagram \citep{M25}. Future models that include LBV phase, binary interactions, or improved treatments of massive star physics may provide opportunities to revisit and refine these analyses. In the subsequent section, we utilize MIST isochrones, with rotation included, to age-date the stellar environment of R127 \& R128 clusters and evaluate its implications for the LBV.  Given the limitations described above, however, we do not draw conclusions from the inferred individual age of the LBV.

\subsection{Sibling Probability}\label{sec:Sibling} 

R127 \& R128, as young open clusters, contain stars that are formed together and thus share common properties, such as age. In this analysis, the age uncertainty is approximately 20\%, corresponding to 2 Myr for a 10 Myr old star.  To investigate whether a selected group of stars could plausibly belong to a coeval sibling population, we apply a probabilistic approach based on their inferred age distributions. The analysis begins by selecting the `n' brightest stars and evaluating, across the Markov Chain Monte Carlo (MCMC) posterior samples, how many of these stars fall into each age bin. These counts are normalized and discretized to assess the collective behavior of the group. MCMC samples in which the total number of stars in any age bin exceeds one are considered consistent with a sibling population. 

Rather than identifying individual stars as siblings, the method tests the broader hypothesis that the selected stars may share a common age. Under this assumption, we retain only the MCMC samples consistent with the sibling criterion described above. This filtering process defines a restricted posterior distribution that reflects the condition of coevality. Stellar parameters such as age and mass are then estimated using only this subset of samples, enabling us to derive properties under the sibling hypothesis.

\section{Results}\label{sec:result} 
\subsection{Most Likely Age}
In this work, we use the \textit{Stellar Ages} \citep[see Section~\ref{sec:method};][]{guzman2025a,M25} algorithm with MIST isochrones including rotation to infer stellar ages in the R127 \& R128 clusters. In the Appendix~\ref{sec:appendix}, we show that consistent results are obtained when using MIST and PARSEC isochrones without inferring rotation, most likely because the isochrones from different models appear very similar during the main-sequence phase.

\begin{figure}
\includegraphics[width=\linewidth, trim=0 0 0 0, clip]{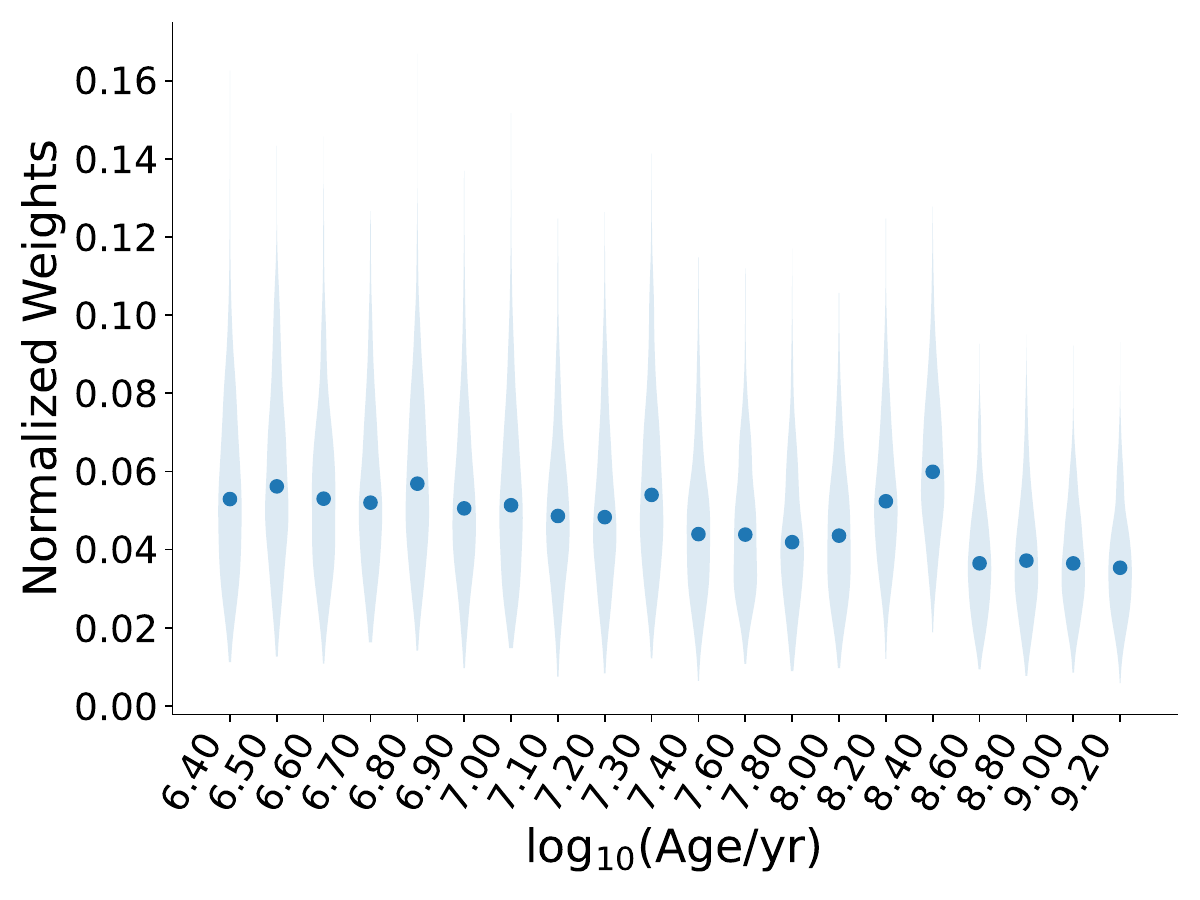}
\caption{Normalized weights as a function of log age marginalized over metallicity and rotation. The central points represent the median age, while the width of the violin indicates the distribution of possible weights. Although there are minor peaks, no strong peak emerges. This underscores the value of a method that provides age constraints on individual stars, especially in cases where the population lacks a clear age signature.
\label{fig:WeightViolin}}
\end{figure}

Fig.~\ref{fig:WeightViolin} shows the inferred weights marginalized over metallicity and rotation. Although several minor peaks are present, for example, at log$_{10}$(t/yr) $\sim$ 6.50 and 6.90, none are prominent, indicating that the available dataset does not yield a clear age for the overall stellar population in this region. This broad range of ages is the first indication that the model does not fit the data.  Furthermore, if one were to interpret these results literally, then this range of ages would imply a wide range of masses (10--100 M$_{\odot}$), all at the most evolved state.  Given the dynamical cohesiveness of the cluster, this spread in ages and masses is more likely an indicator of internal inconsistencies within the modeling framework rather than a true physical distribution. This analysis includes only the population-level weighting constraints and does not incorporate any additional prior assumptions about sibling probability.

\begin{figure}
\includegraphics[width=\linewidth, trim=0 0 0 0, clip]{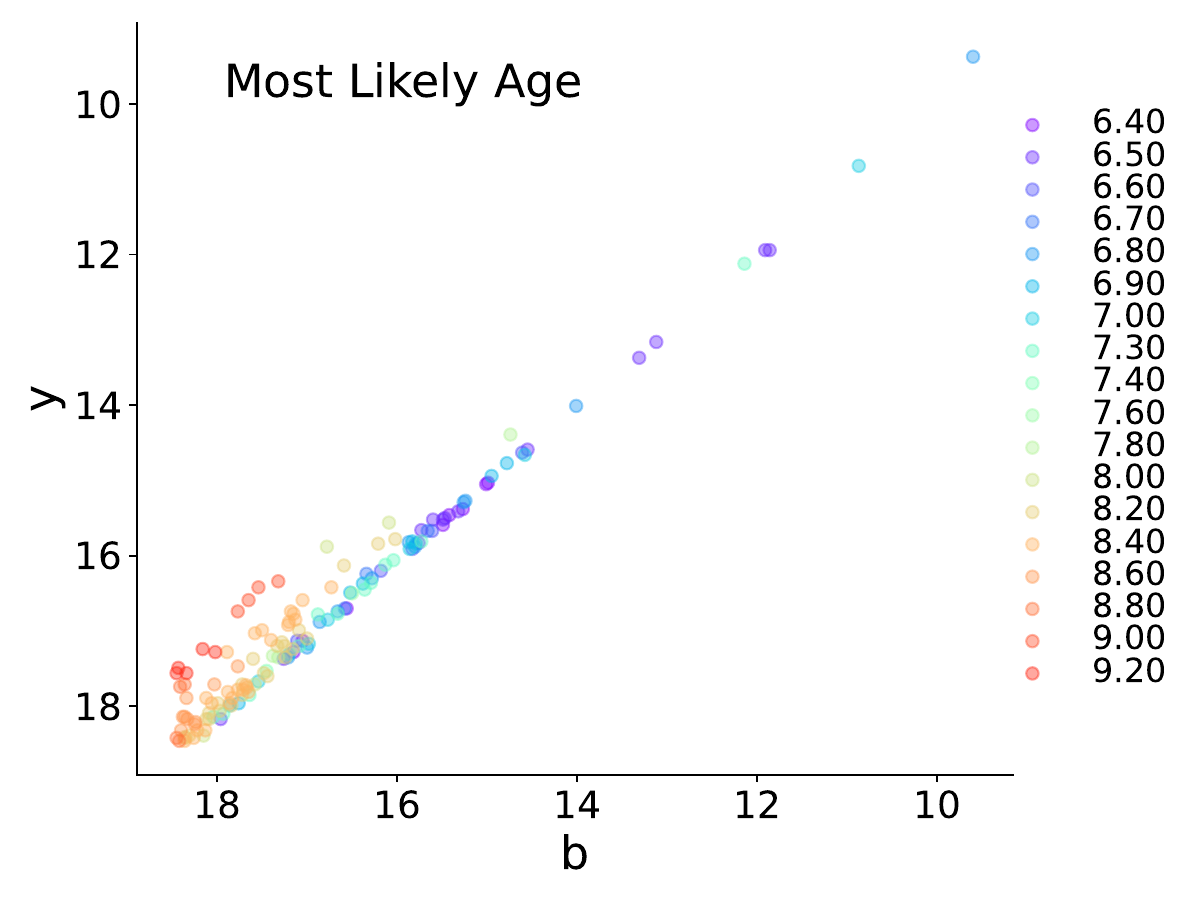}
\caption{The most likely ages inferred for individual stars. The two brightest stars, R127 and R128, have estimated ages of log${10}$(t/yr)$\sim$ 6.8-7.0, while the other bright cluster members are around log${10}$(t/yr)$\sim$ 6.4 old. 
\label{fig:MostLikelyAge}}
\end{figure}

\begin{figure}
\includegraphics[width=\linewidth, trim=0 0 0 0, clip]{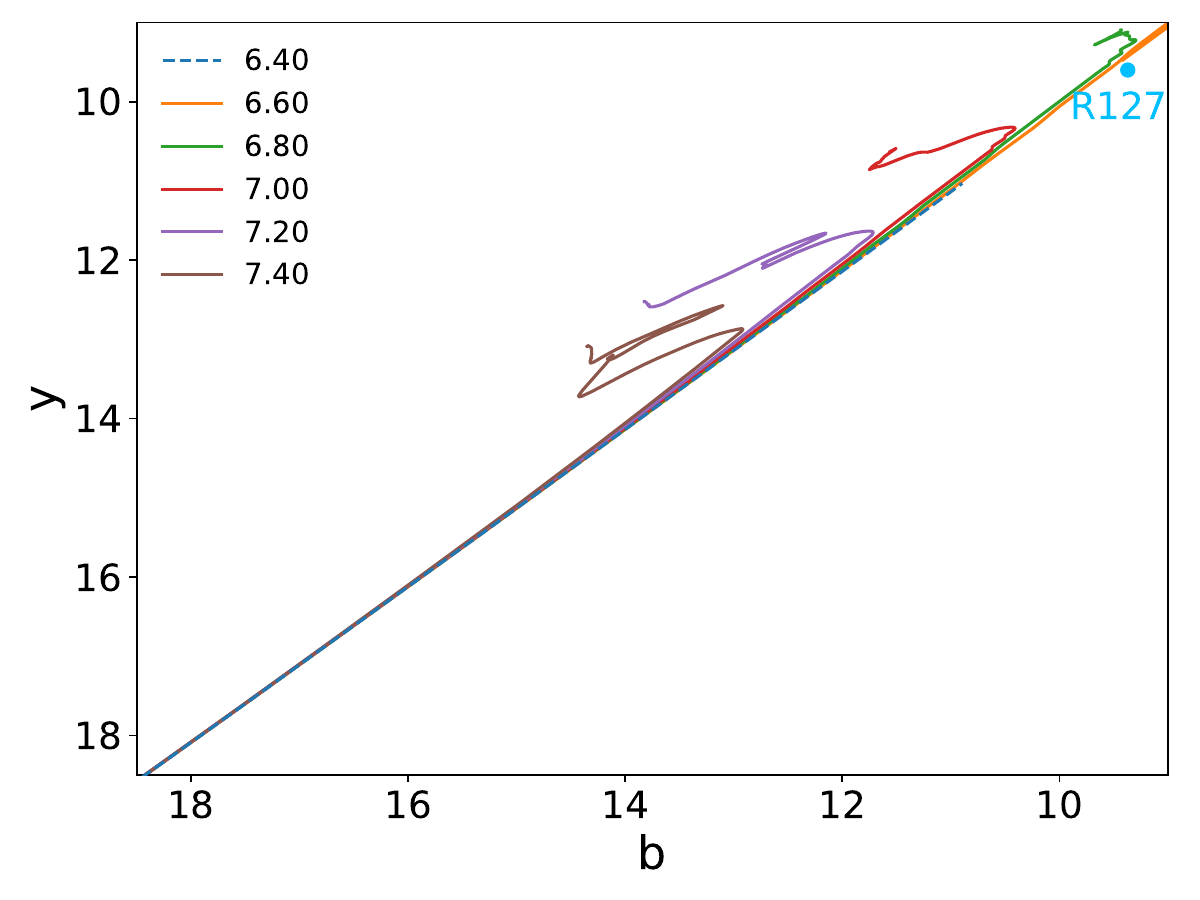}
\caption{Same as Fig.~\ref{fig:models} (first panel), but shown in magnitude space. This figure demonstrates that some evolutionary tracks, such as the 6.4 model, do not extend to the location of R127.
\label{fig:Isoredvsblue_MIST}}
\end{figure}

\begin{figure}
\includegraphics[width=\linewidth, trim=0 0 0 0, clip]{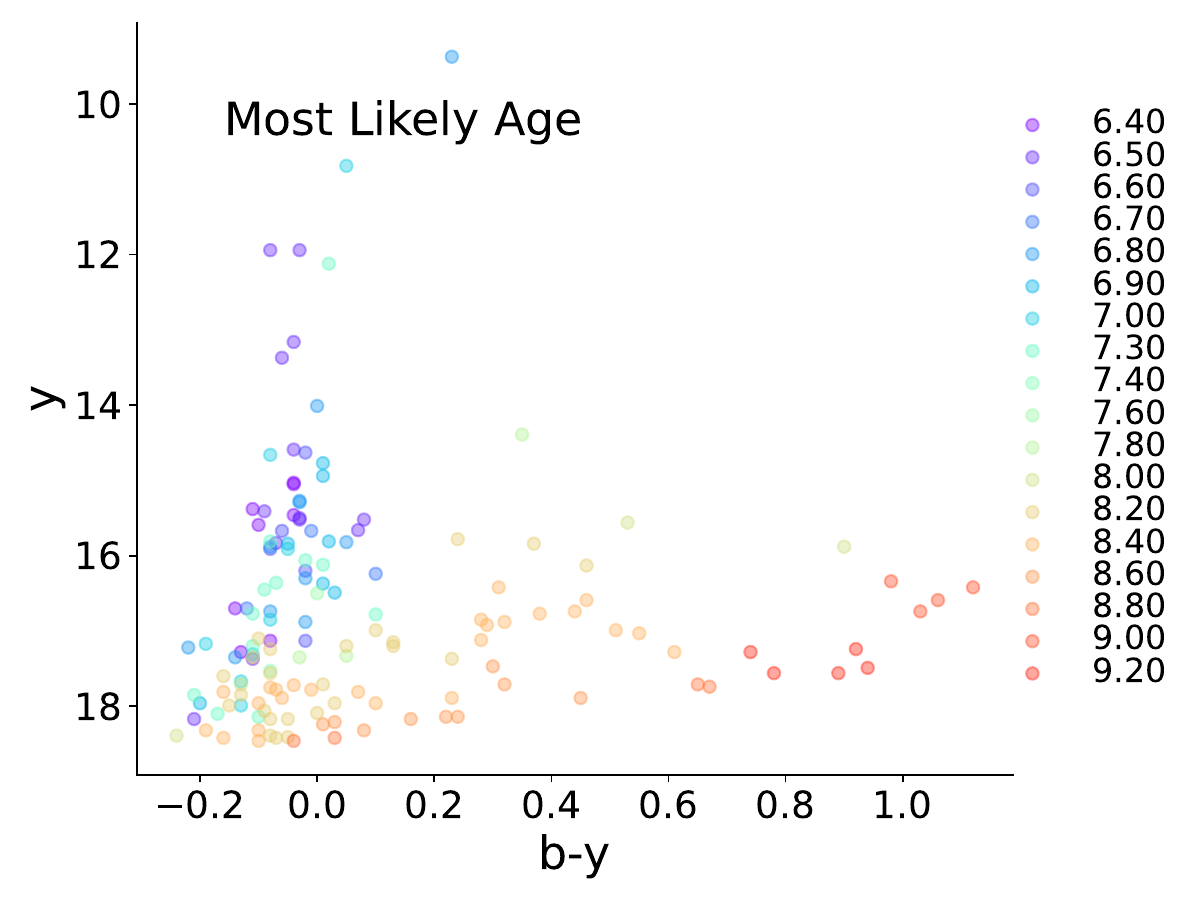}
\caption{Same as Fig.~\ref{fig:MostLikelyAge}, but with color on the x-axis. This highlights that R127 and R128 are slightly redder than the other bright stars.
\label{fig:CMD}}
\end{figure}

The \textit{Stellar Ages} algorithm also provides individual age estimates for each star by inferring the most likely age from its photometry while accounting for the population-level age distribution. Fig.~\ref{fig:MostLikelyAge} shows the most likely ages inferred for all 147 stars within 1 arcminute of the R127 \& R128 clusters. The two brightest members, R127 and R128, have individual age estimates of log$_{10}$(t/yr) $\sim$ 6.8 and 7.0 respectively, while several other bright stars in the region exhibit younger ages around log$_{10}$(t/yr) $\sim$ 6.4. This discrepancy may arise from limitations in the single-star isochrones, as the log$_{10}$(t/yr) $\sim$ 6.4 model does not extend to the locations of the brightest stars, as shown in Fig.~\ref{fig:Isoredvsblue_MIST}. Although the 6.80 model does reach those magnitudes, it is assigned relatively low weight because the other bright stars are not consistent with that age. 
Alternatively, the older ages might be due to R127 and R128 being redder than the other bright stars (see Fig.~\ref{fig:CMD}). To test this, we shifted R128 by 0.3~mag (consistent with the variability amplitude reported by \citet{V98}) and R127 by 0.6~mag toward bluer colors. With this adjustment, R128 is assigned an age of log$_{10}$(t/yr) $\approx$ 6.4, whereas R127 still does not yield an age of 6.4. This indicates that the older inferred age is not a consequence of its redder color, but instead reflects the limitations in the current single-star evolution models to accurately reproduce the effective temperature and luminosity of LBVs. Since single-star models do not include the LBV phase and in some cases do not extend to this region of the HR diagram (see Fig.~\ref{fig:models}), we refrain from drawing conclusions about the individual age of the LBV. Nevertheless, we report the age so that future studies with improved stellar evolution models can compare their predictions directly with our results.

\subsection{The five brightest stars}

Given the lack of a clear age for the full population, we perform a test focused only on the brightest members. In this analysis, we consider the five brightest stars and evaluate what age they imply under the assumption of single-star evolutionary models, temporarily ignoring the lower-mass population. This approach allows us to test whether the brightest stars alone define a younger, more consistent age. It is important to emphasize that if these sources have undergone binary interaction, their inferred single-star ages may not reflect their actual evolutionary histories. The results will be compared with expectations for lower-mass main-sequence stars in the next section.

\begin{figure}
\includegraphics[width=\linewidth, trim=0 0 0 0, clip]{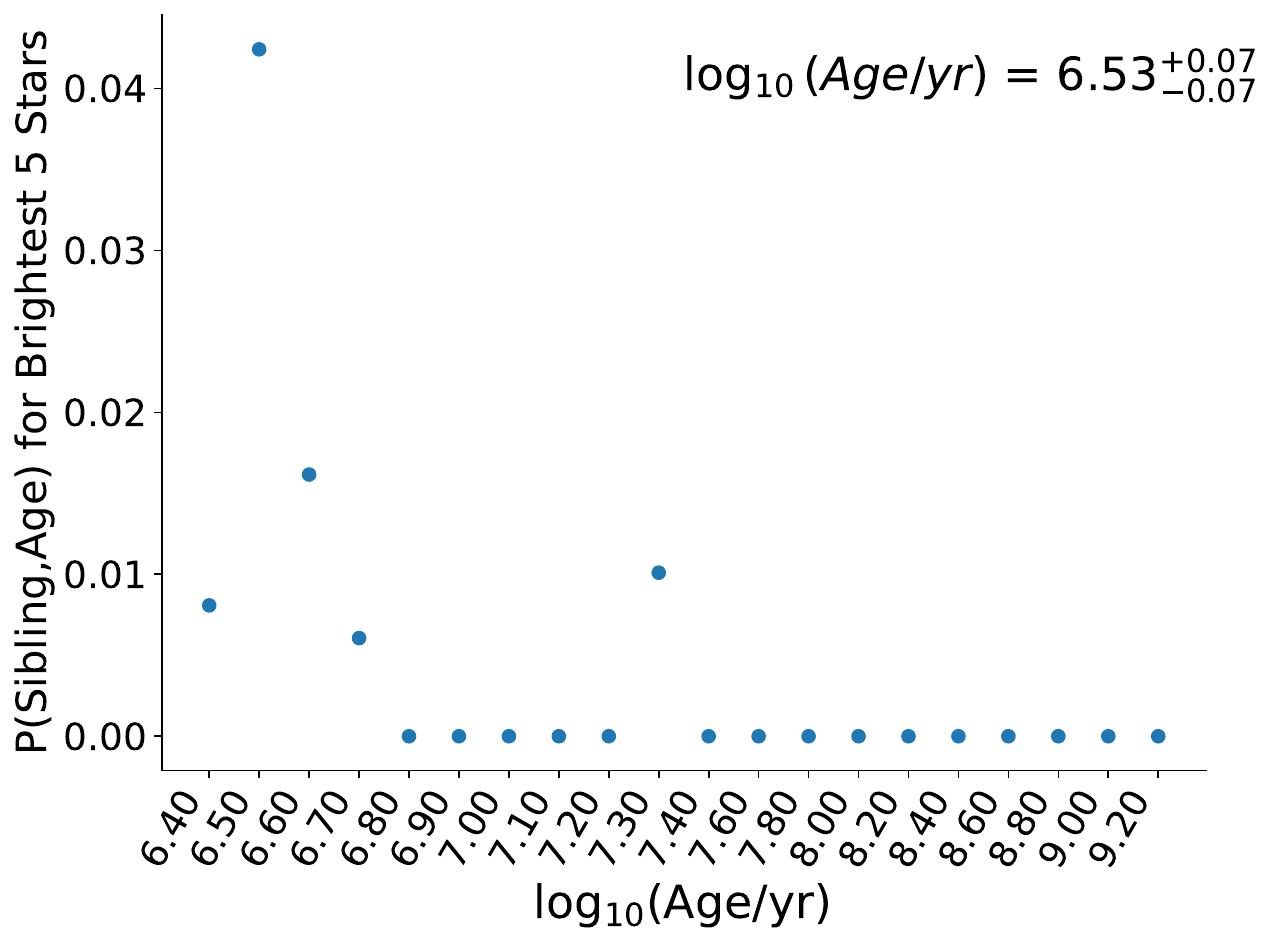}
\caption{The posterior probability distribution of stellar ages for the five brightest stars, based on MCMC draws where they are treated as coeval siblings. Unlike the full stellar population shown in Fig.~\ref{fig:WeightViolin}, which exhibits no clear age, the brightest stars display a pronounced peak at $\log_{10}(t/\mathrm{yr}) = 6.53$. This suggests that their distribution along the main sequence deviates from expectations based on single-star evolution models.
\label{fig:ProbablitySiblings}}
\end{figure}

We therefore refine the cluster age estimate by using the five brightest stars to compute the cumulative probability distribution of stellar ages, under the condition that at least half of them are coeval siblings. 
The resulting probability distribution, shown in Fig.~\ref{fig:ProbablitySiblings}, exhibits a well-defined peak at $\log_{10}(t/\mathrm{yr}) = 6.53 \pm 0.07$. 
This peak is consistent with one of the minor peaks found in Fig.~\ref{fig:WeightViolin} but becomes more statistically significant when restricted to the brightest members. In contrast, the whole population does not exhibit a clear or consistent age distribution because the brightness distribution is inconsistent
with standard population modeling assumptions (see Section~\ref{sec:inconsistencies} for details).

\subsection{Population-level inconsistencies with single-star evolutionary models}\label{sec:inconsistencies}
A critical result is the apparent deficit of low-mass main-sequence stars, relative to expectations from single-star evolutionary models. To illustrate this discrepancy, we define two regions of the magnitude-magnitude diagram used in the comparison. Region A contains the five brightest stars, which dominate the age information for the youngest population. Region B samples the lower-mass main-sequence stars and, based on the available data, is expected to be complete over the magnitude range considered, with an estimated completeness magnitude limit of $\sim$18.

\begin{figure}
\includegraphics[width=\linewidth, trim=0 0 0 0, clip]{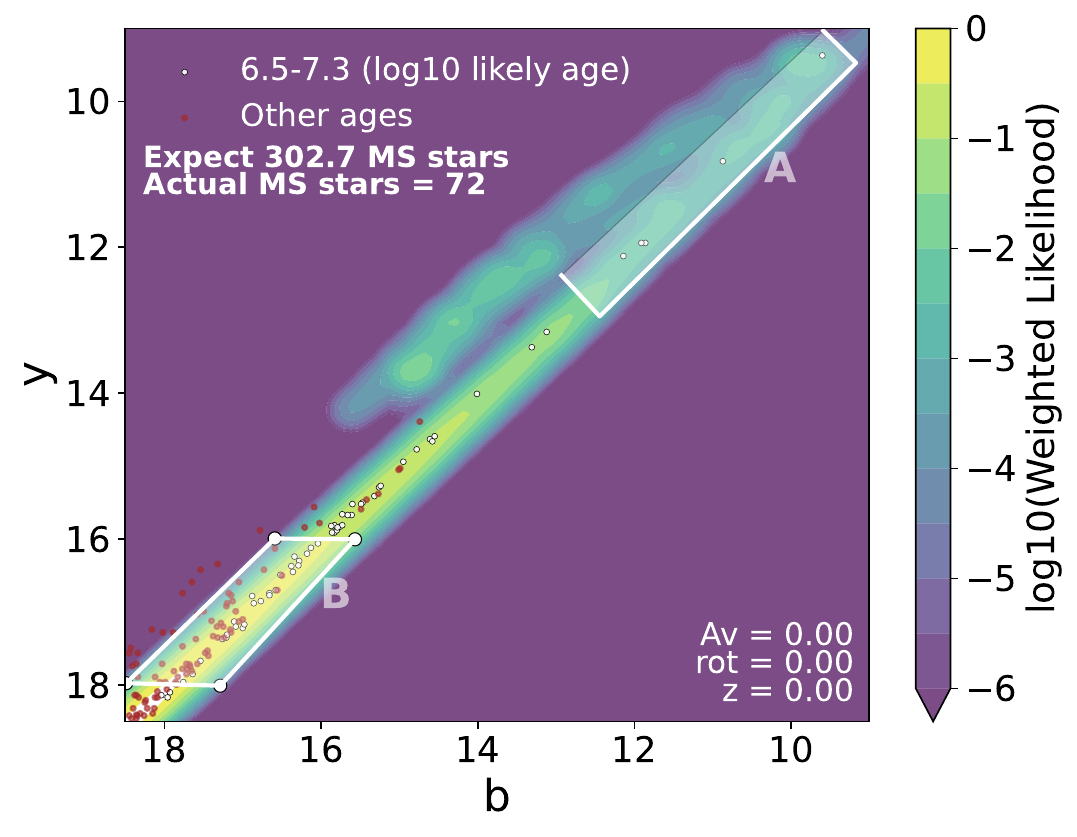}
\caption{Comparison between the observed and expected number of main-sequence stars. White dots indicate stars with inferred ages of log$_{10}$(t/yr) $\sim$ 6.5–7.3, while red dots represent stars of other ages. The background color map shows the age-weighted model likelihoods for the 6.5-7.3 age range. Based on the single-star models and the five brightest stars in region A, the expected number of main-sequence stars in region B is 302.7, compared to the 72 observed. This discrepancy is quantified in Fig.~\ref{fig:posterior_analysis} and indicates tension between the observed and expected values, which may be due to data incompleteness or the bright sources being products of binary evolution or very rapid rotation.
\label{fig:ExpectedNumber}}
\end{figure}

\begin{figure}
\includegraphics[width=\linewidth, trim=0 0 0 0, clip]{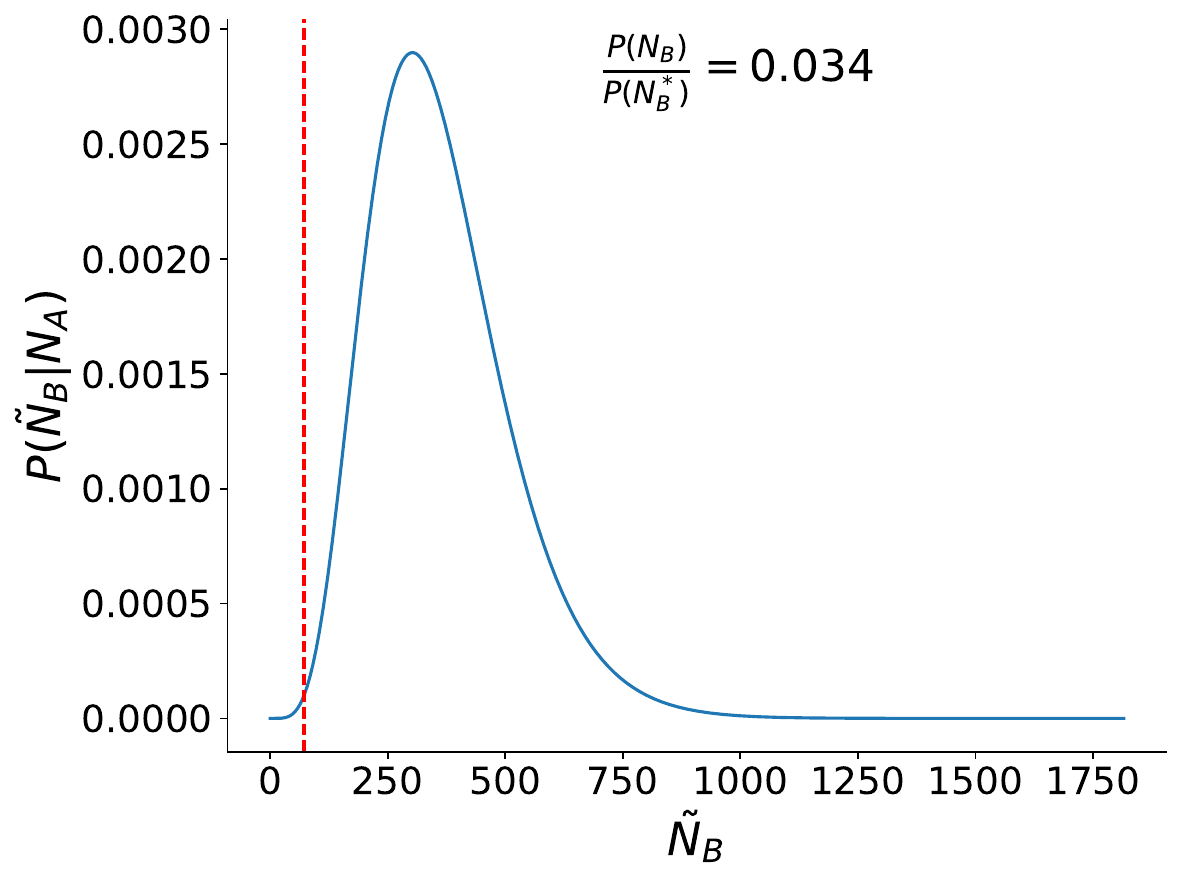}
\caption{Posterior distribution for the expected number of main-sequence stars in region B given the number of stars in region A. The red, dashed, vertical line marks the observed number of main-sequence stars. Although the posterior distribution is broad because of small number statistics, the observed count lies in the low-probability tail. The corresponding Poisson probability
is 0.034, meaning the observed number is much less supported than the model prediction, though not entirely ruled out.
\label{fig:posterior_analysis}}
\end{figure}

Assuming a standard Salpeter initial mass function (IMF), Fig.~\ref{fig:ExpectedNumber} shows that, based on the models and the five brightest stars (region A), the expected number of main-sequence stars in region B is 302.7, whereas only 72 are observed. We get a consistent result when adopting an LMC-like metallicity of [M/H] = $-0.40$. While 302 is far larger than 72, Fig.\ref{fig:posterior_analysis} demonstrates that the posterior distribution is broad because small number statistics from region A are projected to region B. Thus, although the single-star MIST model does not strictly rule out the observed value of 72, it is substantially less likely than the model’s preferred value of 302. Similar mismatches between the observed and expected numbers of stars are also seen in other young clusters in the
Local Group \citep{guzman2025a,guzman2025b,M25}.  The low-mass population is either missing due to incompleteness in the dataset or is genuinely absent due to the bright cluster members being products of binary evolution, such as stellar mergers or mass gainers, or due to very rapid rotation (see \cite{M25} for a detailed discussion ruling out alternative scenarios).  

\begin{figure}
    \centering
    \includegraphics[width=\columnwidth, trim=0 0 0 0, clip]{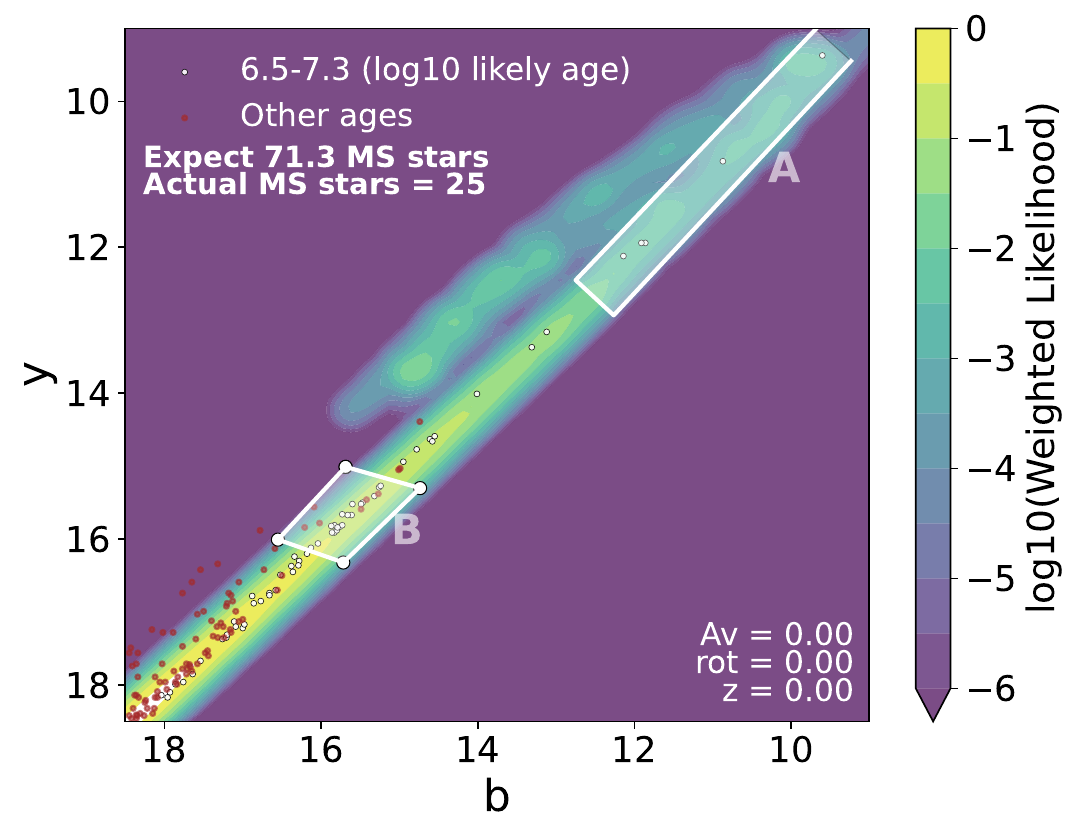}
    \includegraphics[width=\columnwidth, trim=0 0 0 0, clip]{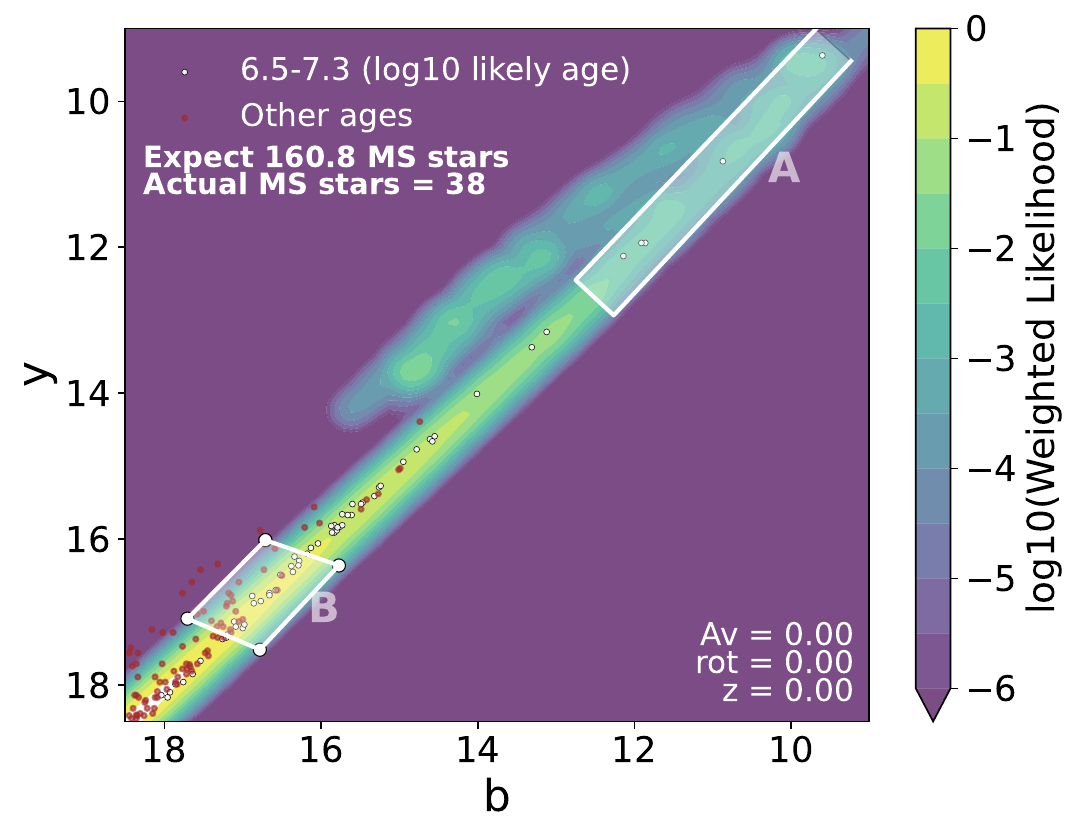}
    \includegraphics[width=\columnwidth, trim=0 0 0 0, clip]{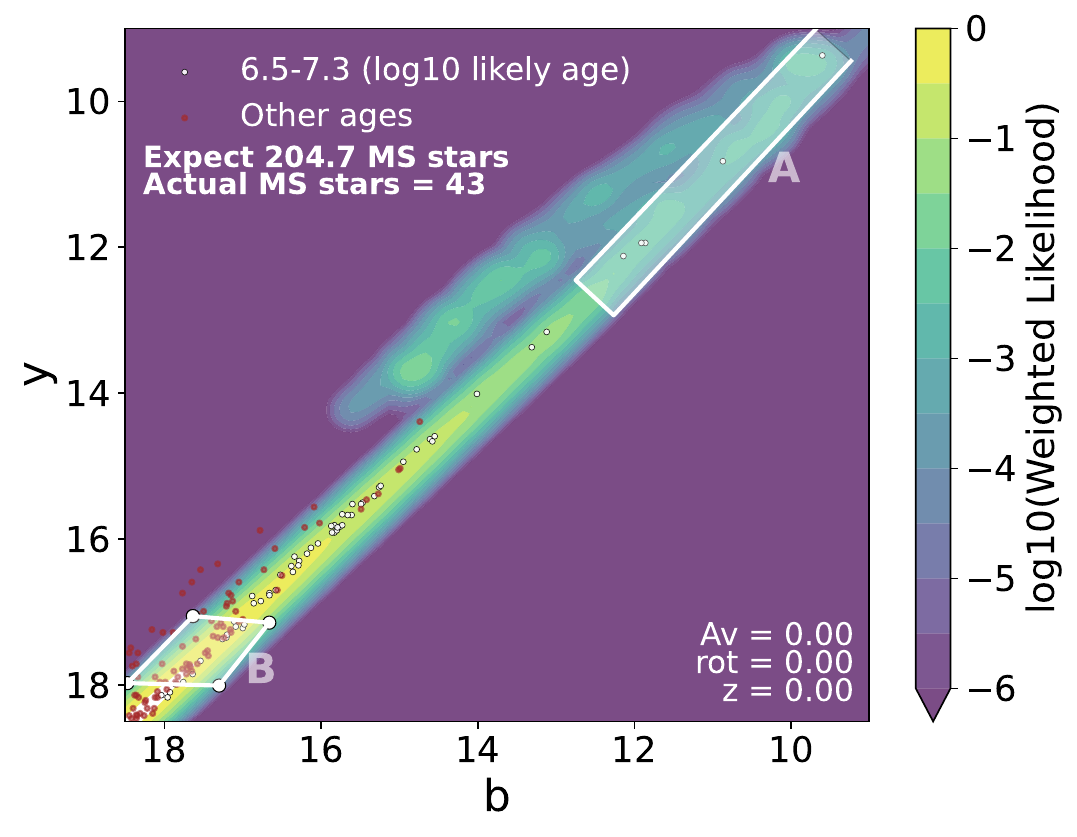}
    \caption{Same as Fig.~\ref{fig:ExpectedNumber}, but for three magnitude bins. The increasing discrepancy between the observed and expected numbers toward fainter magnitudes suggests that the data are likely incomplete.}
    \label{fig:magbin}
\end{figure}

As described in Section~\ref{sec:obs}, the data are most likely complete down to magnitude 18. However, if the completeness limit has been overestimated and the data in region B are in fact incomplete, then the discrepancy between the expected and observed star counts should become even more pronounced at fainter magnitudes. To test this, we repeat the analysis in magnitude bins, and indeed find that the ratio increases toward fainter magnitudes; see Fig.~\ref{fig:magbin}. Although the two faintest bins are consistent within uncertainties, the overall trend is toward a larger discrepancy at lower magnitudes.  

This suggests that the data in region B would need to be incomplete at the level of $\sim$76\%. While some incompleteness is expected ($\lesssim 30\%$; see Section~\ref{sec:obs}), this level substantially exceeds the global deficit in the catalog and would require an implausibly large fraction of stars in this magnitude range to be missed due to crowding and saturation. Incompleteness alone is therefore unlikely to account for the full discrepancy.

\begin{figure*}
    \centering
    \includegraphics[width=0.495\textwidth, trim=0 0 0 0, clip]{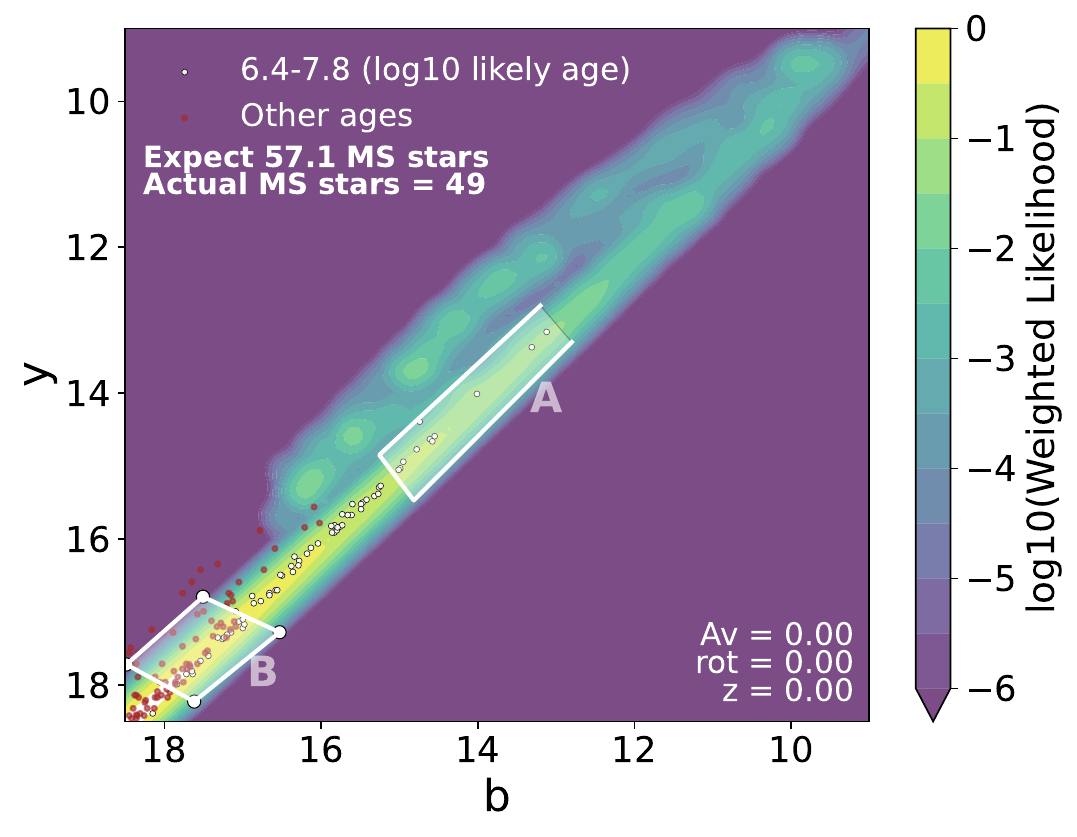}
    \hfill
    \includegraphics[width=0.495\textwidth,trim=0 0 0 0, clip]{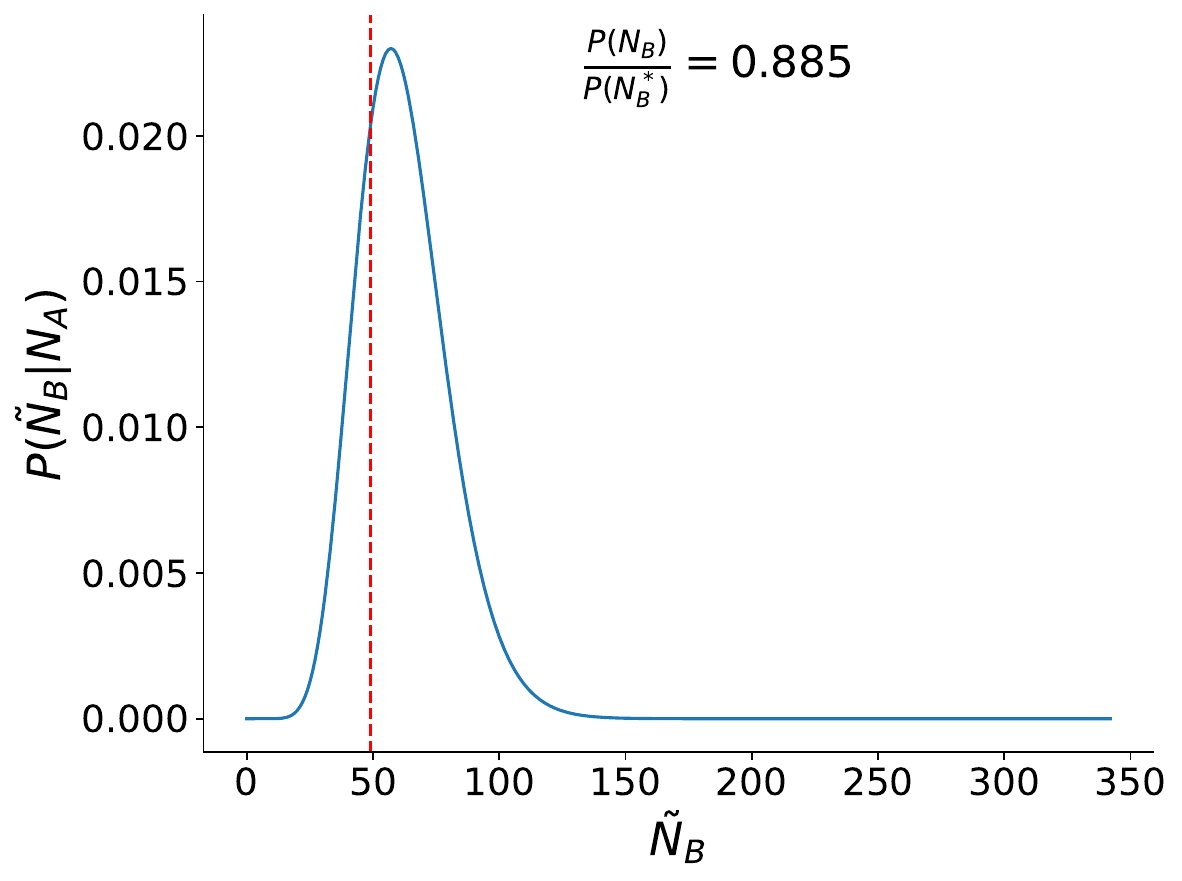}
    \caption{Same as Fig.~\ref{fig:ExpectedNumber} and Fig.~\ref{fig:posterior_analysis}, but with the five brightest stars excluded. The expected and observed numbers are consistent, suggesting that the brightest stars are peculiar relative to the rest of the population.}
    \label{fig:nwithout5}
\end{figure*}

To further assess whether data incompleteness affects the result that the five brightest stars are peculiar, we repeat the analysis after excluding them from the sample. Fig.~\ref{fig:nwithout5} shows that in this case, the expected and observed numbers of stars are more consistent, reinforcing the conclusion that the brightest stars are odd. In addition, when the five brightest stars are excluded, the marginalized posterior distribution over age does not exhibit a statistically significant peak. Thus, the lower-mass population alone does not yield a well-constrained cluster age under single-star evolutionary assumptions.

\begin{figure*}
    \centering
    \includegraphics[width=0.98\columnwidth, trim=10 10 5 10, clip]{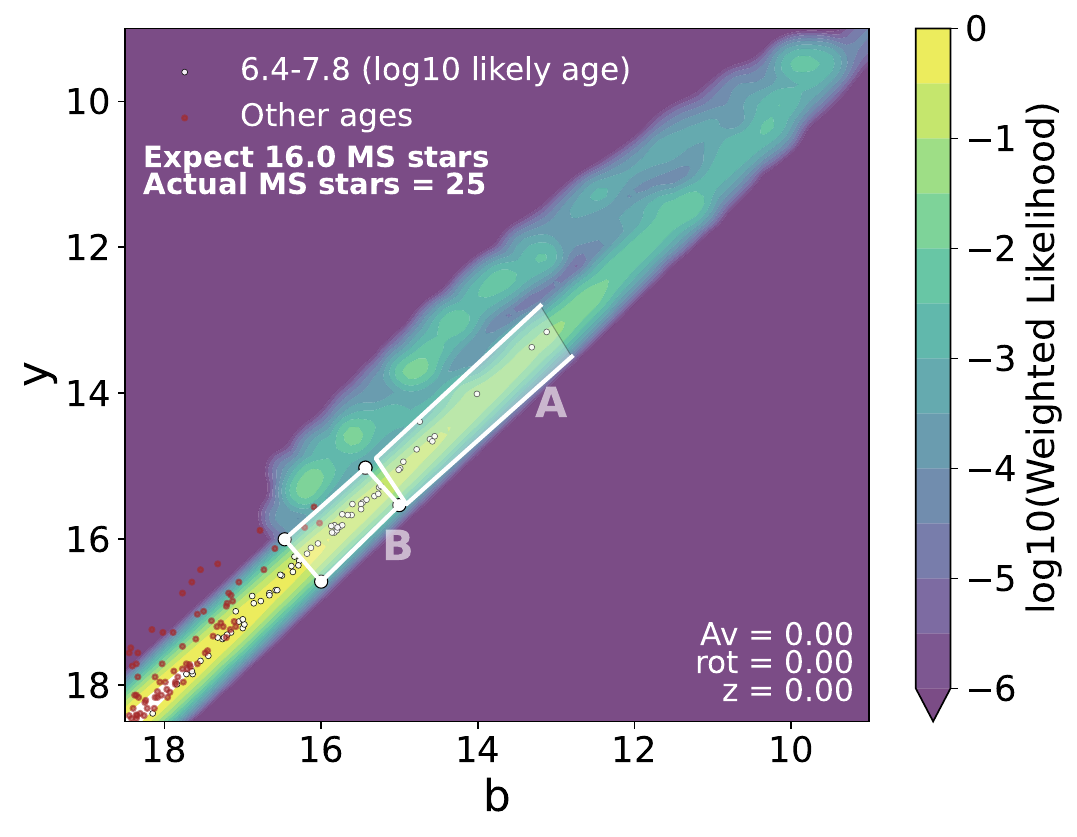}
    \includegraphics[width=0.98\columnwidth, trim=10 10 5 10, clip]{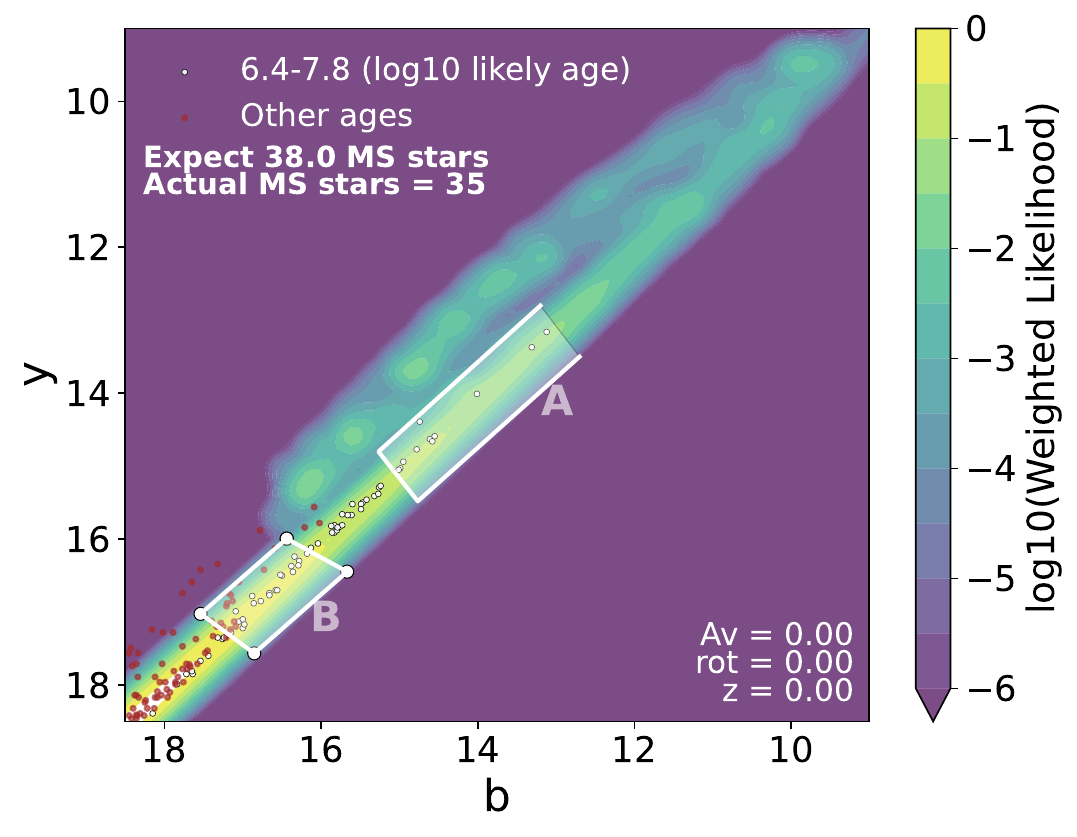}
    \includegraphics[width=0.98\columnwidth, trim=10 10 5 10, clip]{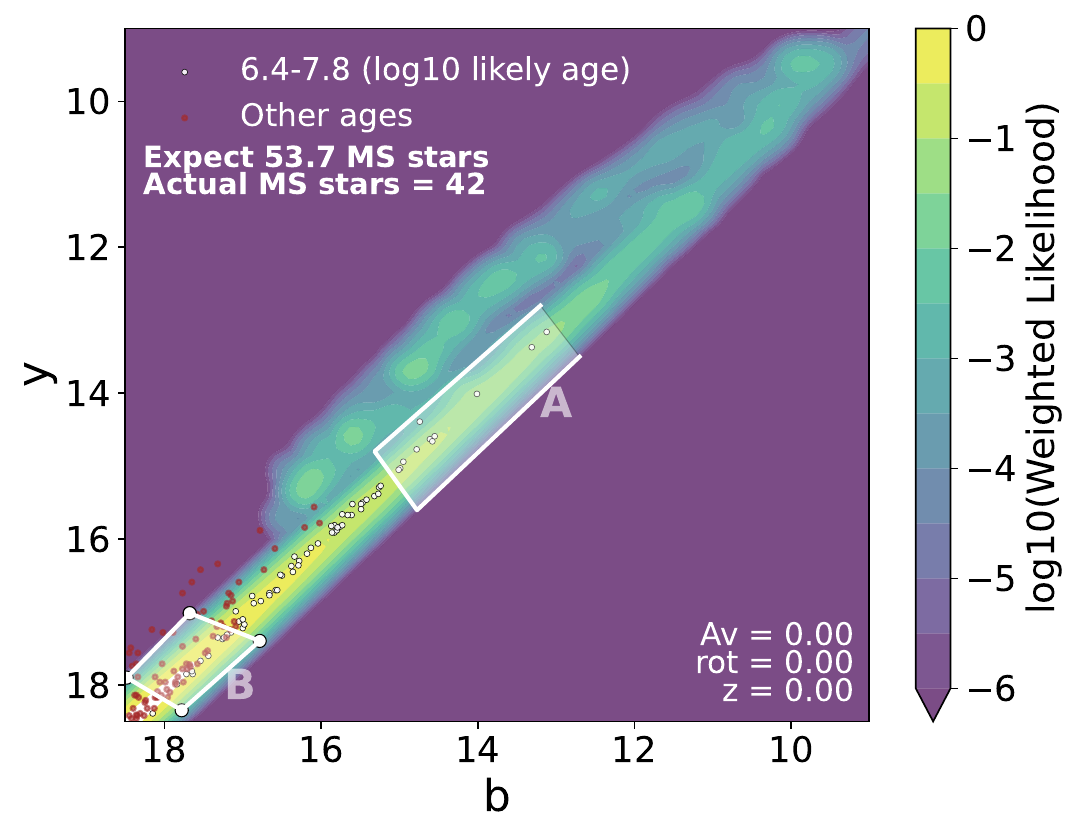}
    
    \caption{Same as Fig.~\ref{fig:magbin}, but with the five brightest stars excluded. The discrepancy between the expected and observed numbers does not increase toward fainter magnitudes, indicating that although the full dataset (including the five brightest stars) is likely incomplete (see Fig.~\ref{fig:magbin}), this incompleteness does not affect the overall conclusion that the brightest members are products of binary evolution or very rapid rotation.}
    \label{fig:magbinwithout5}
\end{figure*}

We further repeat the analysis in three magnitude bins and find that the discrepancy between expected and observed counts does not worsen toward fainter magnitudes; see Fig.~\ref{fig:magbinwithout5}. Taken together, these tests do not exclude the presence of some marginal incompleteness, but they make the severe level of incompleteness discussed above unlikely. More importantly, the absence of a magnitude-dependent trend after removing the five brightest stars reinforces the conclusion that the five brightest stars are most likely products of binary evolution or very rapid rotation.

Binary interaction and rapid rotation can produce overluminous stars within an otherwise coeval population. In close binaries, mass transfer can substantially increase the mass of the accreting secondary and spin it up to high rotation rates; this rotationally enhanced mixing can prolong core hydrogen burning and cause the accretor to appear younger than the bulk of a coeval population using single-star evolutionary tracks \citep{deMink2013}.
Stellar mergers provide a similar pathway, combining two intermediate-mass stars into a single rejuvenated object that can mimic a younger isochrone \citep{Schneider2015}. Population-synthesis models that include binary evolution demonstrate that such interaction products can affect age inferences when single-star models are applied, producing stars that appear younger than the population \citep{Eldridge2017,Stevance2020}. Rapid rotation offers an additional mechanism: rotational mixing transports fresh hydrogen into the core and prolongs core hydrogen burning, leading to hotter and more luminous stars \citep{MeynetMaeder2000,Brott2011}. Moreover, rotation introduces gravity darkening, which causes the observed luminosity and effective temperature to depend on viewing inclination. Together, these effects may provide an explanation for the population-level inconsistency, in which the brightest stars in region A appear younger than the lower-mass main-sequence population in region B when interpreted with single-star evolutionary models and a Salpeter IMF.

\section{Summary} \label{sec:summary}
We constrain the age of R127 by analyzing its local stellar population, the associated R127 \& R128 clusters. R127 is the only confirmed LBV in the LMC that resides within a young star cluster. Its low total velocity relative to the local environment strongly supports that R127 {has not moved far from its birth site and therefore is a member of the R127 cluster \citep{M09,A22,D24}. This makes R127 an ideal case to investigate whether its properties are consistent with a single-star origin or binary evolution.

We utilize Strömgren photometry from the literature \citep{H03} and apply the age-dating algorithm, \textit{Stellar Ages} \citep{guzman2025a,M25}, to a 1-arcminute region around R127. This algorithm infers stellar ages within a population-level Bayesian framework, such that individual age estimates are derived in the context of the full stellar sample rather than in isolation.

Adopting MIST single-star evolutionary tracks with rotation and assuming a Salpeter IMF, we find that the five brightest stars yield a well-defined age of $\log_{10}(t/\mathrm{yr}) = 6.53$. However, a Salpeter mass function extrapolated from these five brightest stars to lower masses would also predict that the number of stars in fainter-magnitude bins on the main sequence should be about 4 times higher than observed.  It is doubtful that all of the missing stars can be attributed to observational incompleteness.  Additionally, the population as a whole lacks a clear age signature because its brightness distribution is inconsistent with expectations from standard population models. Such discrepancies have also been noted in other young clusters across the Local Group \citep{guzman2025a,guzman2025b,M25}, where the data are complete and are typically attributed to binary evolution or very rapid rotation. In our case, incomplete data may also contribute to the inconsistency.

Detailed analysis shows that the discrepancy between the observed and expected number of stars increases toward fainter magnitudes, indicating that the data may be affected by some observational incompleteness. However, when the five brightest stars are excluded, the expected and observed numbers become consistent, and the discrepancy no longer worsens at fainter magnitudes. This suggests that the five brightest stars are indeed peculiar, even though the dataset is still likely incomplete. This challenges the traditional view that LBVs represent a brief transitional phase in the life of single massive stars, but aligns with recent evidence that most massive stars are shaped by binary interaction \citep{S12,M17}.

To further constrain the age of the full population in the R127 \& R128 clusters and confirm whether the bright members are indeed products of binary evolution, a more complete dataset is required. Deep, high–spatial-resolution imaging with the Hubble Space Telescope ({\it HST}) is essential to resolve the crowding in the R127 \& R128 clusters and to detect the faint, low-mass, main-sequence stars. Moreover, for hot massive OB stars and LBVs, the peak of their luminosity lies in the ultraviolet, a wavelength regime that is inaccessible from the ground but critical for accurately characterizing massive stars.  It is also important to extend this analysis to the broader LBV population in the LMC. Unlike R127, all other known LBVs in the LMC appear spatially isolated from massive stars and are not associated with open clusters, which may point to different formation pathways. However, the data required for such an investigation are currently lacking.


\begin{acknowledgments}
Support for M.A. was provided by the VITA-Origins Fellowship, including funding from the Virginia Institute for Theoretical Astrophysics (VITA), supported by the College and Graduate School of Arts and Sciences at the University of Virginia.

\end{acknowledgments}

%





\FloatBarrier
\appendix
\section{Most Likely Ages using Different Models}\label{sec:appendix}
In this Appendix, we present a comparison of the most likely ages inferred using different stellar evolutionary models. Figs.~\ref{fig:MostLikelyAgeParsec} and \ref{fig:MostLikelyAgeMISTwoRotatio} show the results obtained with PARSEC and MIST isochrones (without inferring rotation), respectively. The inferred age distributions are consistent across models, indicating that the choice of isochrones and the inclusion of rotation have only a minor impact on the overall results.

\begin{figure}[h]
    \centering
    \begin{minipage}[t]{0.48\textwidth}
        \centering
        \includegraphics[width=\textwidth]{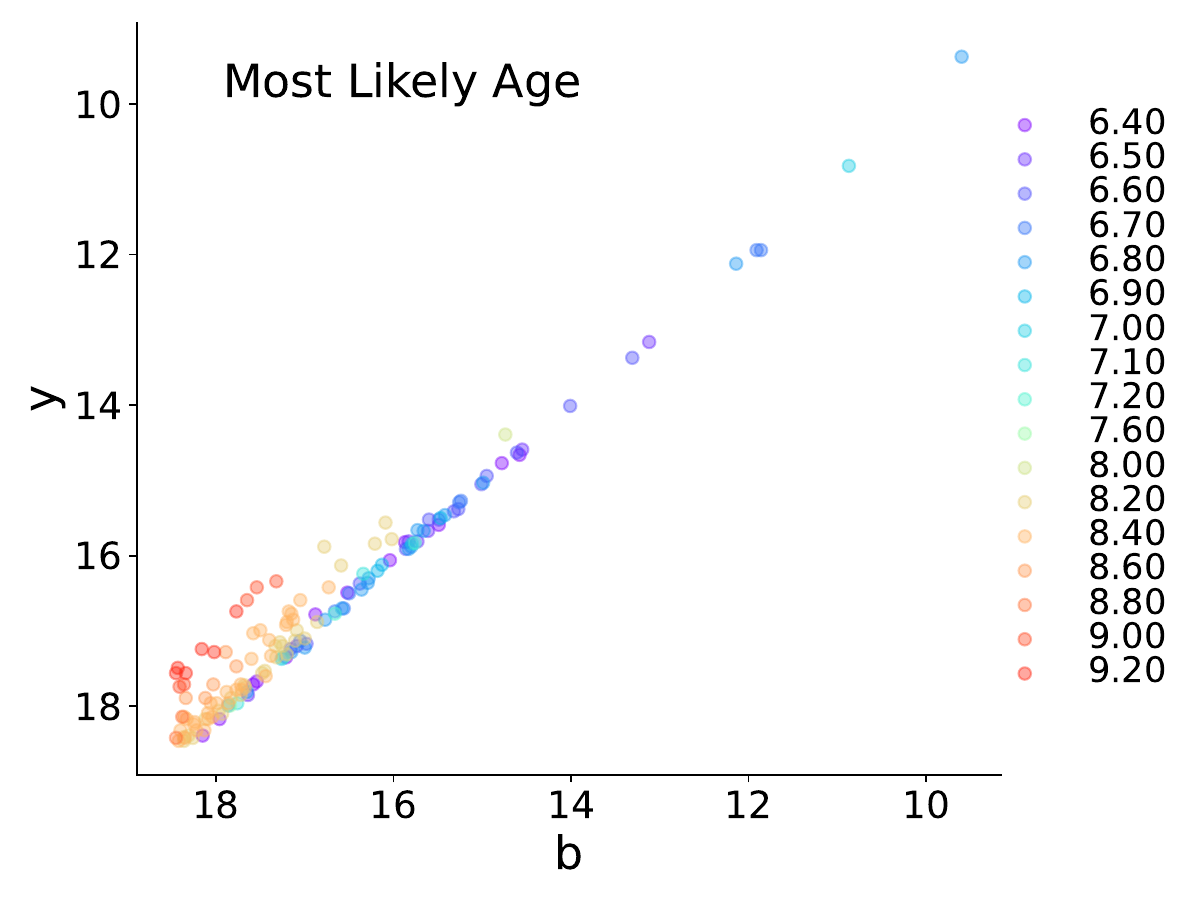}
        \caption{Same as Fig.~\ref{fig:MostLikelyAge}, but using PARSEC models without inferring rotation. 
        The results are consistent across different models.}
        \label{fig:MostLikelyAgeParsec}
    \end{minipage}%
    \hfill
    \begin{minipage}[t]{0.48\textwidth}
        \centering
        \includegraphics[width=\textwidth]{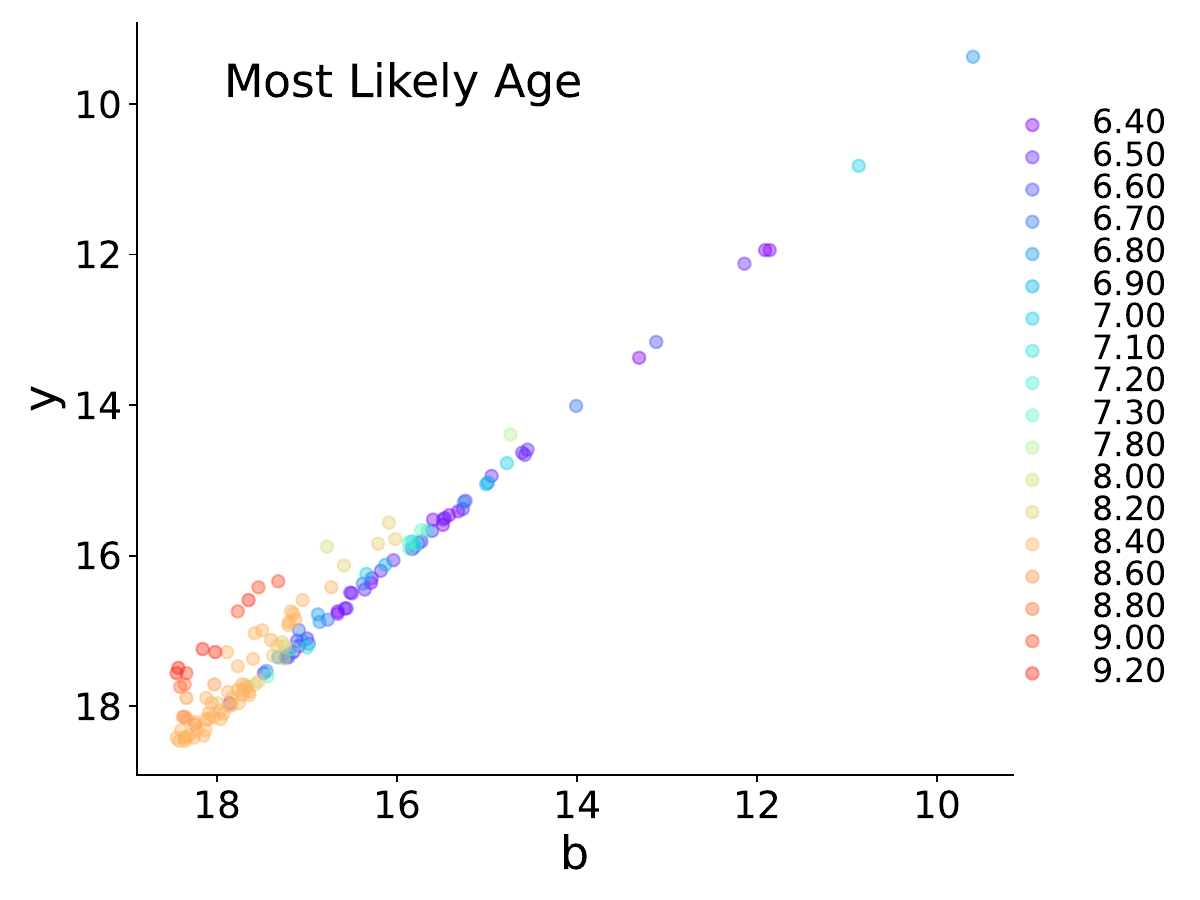}
        \caption{Same as Fig.~\ref{fig:MostLikelyAge}, but using MIST models without inferring rotation. 
        The inclusion of rotation has only a minor impact on the overall stellar distribution.}
        \label{fig:MostLikelyAgeMISTwoRotatio}
    \end{minipage}
\end{figure}


\bibliography{MA}{}

@ARTICLE{S06,
       author = {{Smith}, Nathan and {Owocki}, Stanley P.},
        title = "{On the Role of Continuum-driven Eruptions in the Evolution of Very Massive Stars and Population III Stars}",
      journal = {\apjl},
         year = 2006,
        month = jul,
       volume = {645},
       number = {1},
        pages = {L45-L48},
          doi = {10.1086/506523},
archivePrefix = {arXiv},
       eprint = {astro-ph/0606174},
 primaryClass = {astro-ph},
       adsurl = {https://ui.adsabs.harvard.edu/abs/2006ApJ...645L..45S}
}

@ARTICLE{M00,
       author = {{Massey}, Philip and {Waterhouse}, Elizabeth and {DeGioia-Eastwood}, Kathleen},
        title = "{The Progenitor Masses of Wolf-Rayet Stars and Luminous Blue Variables Determined from Cluster Turnoffs. I. Results from 19 OB Associations in the Magellanic Clouds}",
      journal = {\aj},
         year = 2000,
        month = may,
       volume = {119},
       number = {5},
        pages = {2214-2241},
          doi = {10.1086/301345},
archivePrefix = {arXiv},
       eprint = {astro-ph/0002233},
 primaryClass = {astro-ph},
       adsurl = {https://ui.adsabs.harvard.edu/abs/2000AJ....119.2214M}
}

@ARTICLE{P93,
       author = {{Parker}, Joel Wm. and {Clayton}, Geoffrey C. and {Winge}, Claudia and {Conti}, Peter S.},
        title = "{A New Luminous Blue Variable: R143 in 30 Doradus}",
      journal = {\apj},
         year = 1993,
        month = jun,
       volume = {409},
        pages = {770},
          doi = {10.1086/172706},
       adsurl = {https://ui.adsabs.harvard.edu/abs/1993ApJ...409..770P}
}

@article{guzman2025b,
doi = {10.3847/1538-4357/add265},
url = {https://dx.doi.org/10.3847/1538-4357/add265},
year = {2025b},
month = {jun},
publisher = {The American Astronomical Society},
volume = {986},
number = {1},
pages = {83},
author = {Guzman, Joseph J. and Murphy, Jeremiah W. and Barrientos, Andrés F. and Williams, Benjamin F. and Dalcanton, Julianne J.},
title = {Stellar Ages: A Code to Infer Properties of Stellar Populations},
journal = {The Astrophysical Journal}
}

@ARTICLE{M25,
       author = {{Murphy}, Jeremiah W. and {Barrientos}, Andr{\'e}s F. and {Andrae}, Ren{\'e} and {Guzman}, Joseph and {Williams}, Benjamin F. and {Dalcanton}, Julianne J. and {Koplitz}, Brad},
        title = "{The Mass of the Vela Pulsar Progenitor and the Age of the Vela-Puppis Complex}",
      journal = {\apj},
         year = 2025,
        month = aug,
       volume = {988},
       number = {2},
          eid = {241},
        pages = {241},
          doi = {10.3847/1538-4357/ade5b4},
archivePrefix = {arXiv},
       eprint = {2406.04075},
 primaryClass = {astro-ph.SR},
       adsurl = {https://ui.adsabs.harvard.edu/abs/2025ApJ...988..241M}
}

@ARTICLE{V01,
       author = {{van Genderen}, A.~M.},
        title = "{S Doradus variables in the Galaxy and the Magellanic Clouds}",
      journal = {\aap},
         year = 2001,
        month = feb,
       volume = {366},
        pages = {508-531},
          doi = {10.1051/0004-6361:20000022},
       adsurl = {https://ui.adsabs.harvard.edu/abs/2001A&A...366..508V}
}

@ARTICLE{W89,
       author = {{Wolf}, B.},
        title = "{Empirical amplitude-luminosity relation of S Doradus variables and extragalactic distances.}",
      journal = {\aap},
         year = 1989,
        month = jun,
       volume = {217},
        pages = {87-91},
       adsurl = {https://ui.adsabs.harvard.edu/abs/1989A&A...217...87W}
}

@ARTICLE{B12,
       author = {{Bressan}, Alessandro and {Marigo}, Paola and {Girardi}, L{\'e}o. and {Salasnich}, Bernardo and {Dal Cero}, Claudia and {Rubele}, Stefano and {Nanni}, Ambra},
        title = "{PARSEC: stellar tracks and isochrones with the PAdova and TRieste Stellar Evolution Code}",
      journal = {\mnras},
         year = 2012,
        month = nov,
       volume = {427},
       number = {1},
        pages = {127-145},
          doi = {10.1111/j.1365-2966.2012.21948.x},
archivePrefix = {arXiv},
       eprint = {1208.4498},
 primaryClass = {astro-ph.SR},
       adsurl = {https://ui.adsabs.harvard.edu/abs/2012MNRAS.427..127B}
}

@ARTICLE{C16,
       author = {{Choi}, Jieun and {Dotter}, Aaron and {Conroy}, Charlie and {Cantiello}, Matteo and {Paxton}, Bill and {Johnson}, Benjamin D.},
        title = "{Mesa Isochrones and Stellar Tracks (MIST). I. Solar-scaled Models}",
      journal = {\apj},
         year = 2016,
        month = jun,
       volume = {823},
       number = {2},
          eid = {102},
        pages = {102},
          doi = {10.3847/0004-637X/823/2/102},
archivePrefix = {arXiv},
       eprint = {1604.08592},
 primaryClass = {astro-ph.SR},
       adsurl = {https://ui.adsabs.harvard.edu/abs/2016ApJ...823..102C}
}

@ARTICLE{D16,
       author = {{Dotter}, Aaron},
        title = "{MESA Isochrones and Stellar Tracks (MIST) 0: Methods for the Construction of Stellar Isochrones}",
      journal = {\apjs},
         year = 2016,
        month = jan,
       volume = {222},
       number = {1},
          eid = {8},
        pages = {8},
          doi = {10.3847/0067-0049/222/1/8},
archivePrefix = {arXiv},
       eprint = {1601.05144},
 primaryClass = {astro-ph.SR},
       adsurl = {https://ui.adsabs.harvard.edu/abs/2016ApJS..222....8D}
}

@ARTICLE{G23,
       author = {{Gaia Collaboration} and {Vallenari}, A. and {Brown}, A.~G.~A. and {Prusti}, T. and {de Bruijne}, J.~H.~J. and {Arenou}, F. and {Babusiaux}, C. and {Biermann}, M. and {Creevey}, O.~L. and {Ducourant}, C. and {Evans}, D.~W. and {Eyer}, L. and {Guerra}, R. and {Hutton}, A. and {Jordi}, C. and {Klioner}, S.~A. and {Lammers}, U.~L. and {Lindegren}, L. and {Luri}, X. and {Mignard}, F. and {Panem}, C. and {Pourbaix}, D. and {Randich}, S. and {Sartoretti}, P. and {Soubiran}, C. and {Tanga}, P. and {Walton}, N.~A. and {Bailer-Jones}, C.~A.~L. and {Bastian}, U. and {Drimmel}, R. and {Jansen}, F. and {Katz}, D. and {Lattanzi}, M.~G. and {van Leeuwen}, F. and {Bakker}, J. and {Cacciari}, C. and {Casta{\~n}eda}, J. and {De Angeli}, F. and {Fabricius}, C. and {Fouesneau}, M. and {Fr{\'e}mat}, Y. and {Galluccio}, L. and {Guerrier}, A. and {Heiter}, U. and {Masana}, E. and {Messineo}, R. and {Mowlavi}, N. and {Nicolas}, C. and {Nienartowicz}, K. and {Pailler}, F. and {Panuzzo}, P. and {Riclet}, F. and {Roux}, W. and {Seabroke}, G.~M. and {Sordo}, R. and {Th{\'e}venin}, F. and {Gracia-Abril}, G. and {Portell}, J. and {Teyssier}, D. and {Altmann}, M. and {Andrae}, R. and {Audard}, M. and {Bellas-Velidis}, I. and {Benson}, K. and {Berthier}, J. and {Blomme}, R. and {Burgess}, P.~W. and {Busonero}, D. and {Busso}, G. and {C{\'a}novas}, H. and {Carry}, B. and {Cellino}, A. and {Cheek}, N. and {Clementini}, G. and {Damerdji}, Y. and {Davidson}, M. and {de Teodoro}, P. and {Nu{\~n}ez Campos}, M. and {Delchambre}, L. and {Dell'Oro}, A. and {Esquej}, P. and {Fern{\'a}ndez-Hern{\'a}ndez}, J. and {Fraile}, E. and {Garabato}, D. and {Garc{\'\i}a-Lario}, P. and {Gosset}, E. and {Haigron}, R. and {Halbwachs}, J. -L. and {Hambly}, N.~C. and {Harrison}, D.~L. and {Hern{\'a}ndez}, J. and {Hestroffer}, D. and {Hodgkin}, S.~T. and {Holl}, B. and {Jan{\ss}en}, K. and {Jevardat de Fombelle}, G. and {Jordan}, S. and {Krone-Martins}, A. and {Lanzafame}, A.~C. and {L{\"o}ffler}, W. and {Marchal}, O. and {Marrese}, P.~M. and {Moitinho}, A. and {Muinonen}, K. and {Osborne}, P. and {Pancino}, E. and {Pauwels}, T. and {Recio-Blanco}, A. and {Reyl{\'e}}, C. and {Riello}, M. and {Rimoldini}, L. and {Roegiers}, T. and {Rybizki}, J. and {Sarro}, L.~M. and {Siopis}, C. and {Smith}, M. and {Sozzetti}, A. and {Utrilla}, E. and {van Leeuwen}, M. and {Abbas}, U. and {{\'A}brah{\'a}m}, P. and {Abreu Aramburu}, A. and {Aerts}, C. and {Aguado}, J.~J. and {Ajaj}, M. and {Aldea-Montero}, F. and {Altavilla}, G. and {{\'A}lvarez}, M.~A. and {Alves}, J. and {Anders}, F. and {Anderson}, R.~I. and {Anglada Varela}, E. and {Antoja}, T. and {Baines}, D. and {Baker}, S.~G. and {Balaguer-N{\'u}{\~n}ez}, L. and {Balbinot}, E. and {Balog}, Z. and {Barache}, C. and {Barbato}, D. and {Barros}, M. and {Barstow}, M.~A. and {Bartolom{\'e}}, S. and {Bassilana}, J. -L. and {Bauchet}, N. and {Becciani}, U. and {Bellazzini}, M. and {Berihuete}, A. and {Bernet}, M. and {Bertone}, S. and {Bianchi}, L. and {Binnenfeld}, A. and {Blanco-Cuaresma}, S. and {Blazere}, A. and {Boch}, T. and {Bombrun}, A. and {Bossini}, D. and {Bouquillon}, S. and {Bragaglia}, A. and {Bramante}, L. and {Breedt}, E. and {Bressan}, A. and {Brouillet}, N. and {Brugaletta}, E. and {Bucciarelli}, B. and {Burlacu}, A. and {Butkevich}, A.~G. and {Buzzi}, R. and {Caffau}, E. and {Cancelliere}, R. and {Cantat-Gaudin}, T. and {Carballo}, R. and {Carlucci}, T. and {Carnerero}, M.~I. and {Carrasco}, J.~M. and {Casamiquela}, L. and {Castellani}, M. and {Castro-Ginard}, A. and {Chaoul}, L. and {Charlot}, P. and {Chemin}, L. and {Chiaramida}, V. and {Chiavassa}, A. and {Chornay}, N. and {Comoretto}, G. and {Contursi}, G. and {Cooper}, W.~J. and {Cornez}, T. and {Cowell}, S. and {Crifo}, F. and {Cropper}, M. and {Crosta}, M. and {Crowley}, C. and {Dafonte}, C. and {Dapergolas}, A. and {David}, M. and {David}, P. and {de Laverny}, P. and {De Luise}, F. and {De March}, R.},
        title = "{Gaia Data Release 3. Summary of the content and survey properties}",
      journal = {\aap},
         year = 2023,
        month = jun,
       volume = {674},
          eid = {A1},
        pages = {A1},
          doi = {10.1051/0004-6361/202243940},
archivePrefix = {arXiv},
       eprint = {2208.00211},
 primaryClass = {astro-ph.GA},
       adsurl = {https://ui.adsabs.harvard.edu/abs/2023A&A...674A...1G}
}

@INPROCEEDINGS{C98,
       author = {{Conti}, P.~S.},
        title = "{Basic Observational Constraints on the Evolution of Massive Stars}",
    booktitle = {Observational Tests of the Stellar Evolution Theory},
         year = 1984,
       editor = {{Maeder}, A. and {Renzini}, A.},
       series = {IAU Symposium},
       volume = {105},
        month = jan,
        pages = {233},
       adsurl = {https://ui.adsabs.harvard.edu/abs/1984IAUS..105..233C}
}

@ARTICLE{V98,
       author = {{van Genderen}, A.~M. and {Sterken}, C. and {de Groot}, M. and {Reijns}, R.~A.},
        title = "{Light variations of massive stars (alpha Cyg variables). XV. The LMC supergiants R99 (LBV), R103, R123 (LBV) and R128}",
      journal = {\aap},
         year = 1998,
        month = apr,
       volume = {332},
        pages = {857-866},
       adsurl = {https://ui.adsabs.harvard.edu/abs/1998A&A...332..857V}
}

@ARTICLE{S83,
       author = {{Stahl}, O. and {Wolf}, B. and {Klare}, G. and {Cassatella}, A. and {Krautter}, J. and {Persi}, P. and {Ferrari-Toniolo}, M.},
        title = "{R 127 : an S DOR type variable intermediate between Of and WN.}",
      journal = {\aap},
         year = 1983,
        month = nov,
       volume = {127},
        pages = {49-62},
       adsurl = {https://ui.adsabs.harvard.edu/abs/1983A&A...127...49S}
}

@ARTICLE{D24,
       author = {{Deman}, Julian A. and {Oey}, M.~S.},
        title = "{Kinematic Insights into Luminous Blue Variables and B[e] Supergiants}",
      journal = {\apj},
         year = 2024,
        month = nov,
       volume = {976},
       number = {1},
          eid = {125},
        pages = {125},
          doi = {10.3847/1538-4357/ad813410.1134/S1063772908070019},
archivePrefix = {arXiv},
       eprint = {2410.06448},
 primaryClass = {astro-ph.SR},
       adsurl = {https://ui.adsabs.harvard.edu/abs/2024ApJ...976..125D}
}

@ARTICLE{M09,
       author = {{Munari}, U. and {Siviero}, A. and {Bienaym{\'e}}, O. and {Binney}, J. and {Bland-Hawthorn}, J. and {Campbell}, R. and {Freeman}, K.~C. and {Fulbright}, J.~P. and {Gibson}, B.~K. and {Gilmore}, G. and {Grebel}, E.~K. and {Helmi}, A. and {Navarro}, J.~F. and {Parker}, Q.~A. and {Reid}, W. and {Seabroke}, G.~M. and {Siebert}, A. and {Steinmetz}, M. and {Watson}, F.~G. and {Williams}, M. and {Wyse}, R.~F.~G. and {Zwitter}, T.},
        title = "{RAVE spectroscopy of luminous blue variables in the Large Magellanic Cloud}",
      journal = {\aap},
         year = 2009,
        month = aug,
       volume = {503},
       number = {2},
        pages = {511-520},
          doi = {10.1051/0004-6361/200912398},
archivePrefix = {arXiv},
       eprint = {0907.0177},
 primaryClass = {astro-ph.SR},
       adsurl = {https://ui.adsabs.harvard.edu/abs/2009A&A...503..511M}
}

@ARTICLE{A22,
       author = {{Aghakhanloo}, Mojgan and {Smith}, Nathan and {Andrews}, Jennifer and {Olsen}, Knut and {Besla}, Gurtina and {Choi}, Yumi},
        title = "{Kinematics of luminous blue variables in the Large Magellanic Cloud}",
      journal = {\mnras},
         year = 2022,
        month = oct,
       volume = {516},
       number = {2},
        pages = {2142-2161},
          doi = {10.1093/mnras/stac2265},
archivePrefix = {arXiv},
       eprint = {2202.06887},
 primaryClass = {astro-ph.SR},
       adsurl = {https://ui.adsabs.harvard.edu/abs/2022MNRAS.516.2142A}
}

@ARTICLE{deMink2013,
       author = {{de Mink}, S.~E. and {Langer}, N. and {Izzard}, R.~G. and {Sana}, H. and {de Koter}, A.},
        title = "{The Rotation Rates of Massive Stars: The Role of Binary Interaction through Tides, Mass Transfer, and Mergers}",
      journal = {\apj},
         year = 2013,
        month = feb,
       volume = {764},
       number = {2},
          eid = {166},
        pages = {166},
          doi = {10.1088/0004-637X/764/2/166},
archivePrefix = {arXiv},
       eprint = {1211.3742},
 primaryClass = {astro-ph.SR},
       adsurl = {https://ui.adsabs.harvard.edu/abs/2013ApJ...764..166D}
}

@ARTICLE{Schneider2015,
       author = {{Schneider}, F.~R.~N. and {Izzard}, R.~G. and {Langer}, N. and {de Mink}, S.~E.},
        title = "{Evolution of Mass Functions of Coeval Stars through Wind Mass Loss and Binary Interactions}",
      journal = {\apj},
         year = 2015,
        month = may,
       volume = {805},
       number = {1},
          eid = {20},
        pages = {20},
          doi = {10.1088/0004-637X/805/1/20},
archivePrefix = {arXiv},
       eprint = {1504.01735},
 primaryClass = {astro-ph.SR},
       adsurl = {https://ui.adsabs.harvard.edu/abs/2015ApJ...805...20S}
}

@ARTICLE{D17,
       author = {{De Marco}, Orsola and {Izzard}, Robert G.},
        title = "{Dawes Review 6: The Impact of Companions on Stellar Evolution}",
      journal = {\pasa},
         year = 2017,
        month = jan,
       volume = {34},
          eid = {e001},
        pages = {e001},
          doi = {10.1017/pasa.2016.52},
archivePrefix = {arXiv},
       eprint = {1611.03542},
 primaryClass = {astro-ph.SR},
       adsurl = {https://ui.adsabs.harvard.edu/abs/2017PASA...34....1D}
}

@ARTICLE{V07,
       author = {{Vanbeveren}, D. and {Van Bever}, J. and {Belkus}, H.},
        title = "{The Wolf-Rayet Population Predicted by Massive Single Star and Massive Binary Evolution}",
      journal = {\apjl},
         year = 2007,
        month = jun,
       volume = {662},
       number = {2},
        pages = {L107-L110},
          doi = {10.1086/519454},
archivePrefix = {arXiv},
       eprint = {astro-ph/0703796},
 primaryClass = {astro-ph},
       adsurl = {https://ui.adsabs.harvard.edu/abs/2007ApJ...662L.107V}
}

@ARTICLE{Eldridge2017,
       author = {{Eldridge}, J.~J. and {Stanway}, E.~R. and {Xiao}, L. and {McClelland}, L.~A.~S. and {Taylor}, G. and {Ng}, M. and {Greis}, S.~M.~L. and {Bray}, J.~C.},
        title = "{Binary Population and Spectral Synthesis Version 2.1: Construction, Observational Verification, and New Results}",
      journal = {\pasa},
         year = 2017,
        month = nov,
       volume = {34},
          eid = {e058},
        pages = {e058},
          doi = {10.1017/pasa.2017.51},
archivePrefix = {arXiv},
       eprint = {1710.02154},
 primaryClass = {astro-ph.SR},
       adsurl = {https://ui.adsabs.harvard.edu/abs/2017PASA...34...58E}
}

@ARTICLE{MeynetMaeder2000,
       author = {{Meynet}, G. and {Maeder}, A.},
        title = "{Stellar evolution with rotation. V. Changes in all the outputs of massive star models}",
      journal = {\aap},
         year = 2000,
        month = sep,
       volume = {361},
        pages = {101-120},
          doi = {10.48550/arXiv.astro-ph/0006404},
archivePrefix = {arXiv},
       eprint = {astro-ph/0006404},
 primaryClass = {astro-ph},
       adsurl = {https://ui.adsabs.harvard.edu/abs/2000A&A...361..101M}
}

@ARTICLE{Brott2011,
       author = {{Brott}, I. and {de Mink}, S.~E. and {Cantiello}, M. and {Langer}, N. and {de Koter}, A. and {Evans}, C.~J. and {Hunter}, I. and {Trundle}, C. and {Vink}, J.~S.},
        title = "{Rotating massive main-sequence stars. I. Grids of evolutionary models and isochrones}",
      journal = {\aap},
         year = 2011,
        month = jun,
       volume = {530},
          eid = {A115},
        pages = {A115},
          doi = {10.1051/0004-6361/201016113},
archivePrefix = {arXiv},
       eprint = {1102.0530},
 primaryClass = {astro-ph.SR},
       adsurl = {https://ui.adsabs.harvard.edu/abs/2011A&A...530A.115B}
}

@ARTICLE{Stevance2020,
       author = {{Stevance}, H.~F. and {Eldridge}, J.~J. and {McLeod}, A. and {Stanway}, E.~R. and {Chrimes}, A.~A.},
        title = "{A systematic ageing method I: H II regions D118 and D119 in NGC 300}",
      journal = {\mnras},
         year = 2020,
        month = oct,
       volume = {498},
       number = {1},
        pages = {1347-1363},
          doi = {10.1093/mnras/staa2428},
archivePrefix = {arXiv},
       eprint = {2004.02883},
 primaryClass = {astro-ph.SR},
       adsurl = {https://ui.adsabs.harvard.edu/abs/2020MNRAS.498.1347S}
}

@ARTICLE{S12,
       author = {{Sana}, H. and {de Mink}, S.~E. and {de Koter}, A. and {Langer}, N. and {Evans}, C.~J. and {Gieles}, M. and {Gosset}, E. and {Izzard}, R.~G. and {Le Bouquin}, J. -B. and {Schneider}, F.~R.~N.},
        title = "{Binary Interaction Dominates the Evolution of Massive Stars}",
      journal = {Science},
         year = 2012,
        month = jul,
       volume = {337},
       number = {6093},
        pages = {444},
          doi = {10.1126/science.1223344},
archivePrefix = {arXiv},
       eprint = {1207.6397},
 primaryClass = {astro-ph.SR},
       adsurl = {https://ui.adsabs.harvard.edu/abs/2012Sci...337..444S}
}

@ARTICLE{M17,
       author = {{Moe}, Maxwell and {Di Stefano}, Rosanne},
        title = "{Mind Your Ps and Qs: The Interrelation between Period (P) and Mass-ratio (Q) Distributions of Binary Stars}",
      journal = {\apjs},
         year = 2017,
        month = jun,
       volume = {230},
       number = {2},
          eid = {15},
        pages = {15},
          doi = {10.3847/1538-4365/aa6fb6},
archivePrefix = {arXiv},
       eprint = {1606.05347},
 primaryClass = {astro-ph.SR},
       adsurl = {https://ui.adsabs.harvard.edu/abs/2017ApJS..230...15M}
}

@ARTICLE{A17,
       author = {{Aghakhanloo}, Mojgan and {Murphy}, Jeremiah W. and {Smith}, Nathan and {Hlo{\v{z}}ek}, Ren{\'e}e},
        title = "{Modelling luminous-blue-variable isolation}",
      journal = {\mnras},
         year = 2017,
        month = nov,
       volume = {472},
       number = {1},
        pages = {591-603},
          doi = {10.1093/mnras/stx2050},
archivePrefix = {arXiv},
       eprint = {1701.05626},
 primaryClass = {astro-ph.SR},
       adsurl = {https://ui.adsabs.harvard.edu/abs/2017MNRAS.472..591A}
}

@ARTICLE{S15,
       author = {{Smith}, Nathan and {Tombleson}, Ryan},
        title = "{Luminous blue variables are antisocial: their isolation implies that they are kicked mass gainers in binary evolution}",
      journal = {\mnras},
         year = 2015,
        month = feb,
       volume = {447},
       number = {1},
        pages = {598-617},
          doi = {10.1093/mnras/stu2430},
archivePrefix = {arXiv},
       eprint = {1406.7431},
 primaryClass = {astro-ph.SR},
       adsurl = {https://ui.adsabs.harvard.edu/abs/2015MNRAS.447..598S}
}

@ARTICLE{guzman2025a,
       author = {{Guzman}, Joseph and {Murphy}, Jeremiah and {Beasor}, Emma and {Dalcanton}, Julianne and {Smith}, Nathan and {Aghakhanloo}, Mojgan and {Williams}, Benjamin and {Barrientos}, Andres},
        title = "{Quantifying Systematic Age Discrepancies in Very Young Star Clusters}",
      journal = {arXiv e-prints},
         year = {2025a},
        month = dec,
          eid = {arXiv:2512.17033},
        pages = {arXiv:2512.17033},
          doi = {10.48550/arXiv.2512.17033},
archivePrefix = {arXiv},
       eprint = {2512.17033},
 primaryClass = {astro-ph.SR},
       adsurl = {https://ui.adsabs.harvard.edu/abs/2025arXiv251217033G}
}

@INPROCEEDINGS{S26,
       author = {{Smith}, Nathan},
        title = "{Luminous blue variables}",
    booktitle = {Encyclopedia of Astrophysics, Volume 2},
         year = 2026,
       volume = {2},
        month = jan,
        pages = {508-532},
          doi = {10.1016/B978-0-443-21439-4.00147-4},
archivePrefix = {arXiv},
       eprint = {2509.22990},
 primaryClass = {astro-ph.SR},
       adsurl = {https://ui.adsabs.harvard.edu/abs/2026enap....2..508S}
}

@ARTICLE{J14,
       author = {{Justham}, Stephen and {Podsiadlowski}, Philipp and {Vink}, Jorick S.},
        title = "{Luminous Blue Variables and Superluminous Supernovae from Binary Mergers}",
      journal = {\apj},
         year = 2014,
        month = dec,
       volume = {796},
       number = {2},
          eid = {121},
        pages = {121},
          doi = {10.1088/0004-637X/796/2/121},
archivePrefix = {arXiv},
       eprint = {1410.2426},
 primaryClass = {astro-ph.SR},
       adsurl = {https://ui.adsabs.harvard.edu/abs/2014ApJ...796..121J}
}

@ARTICLE{B22,
       author = {{Brennan}, S.~J. and {Elias-Rosa}, N. and {Fraser}, M. and {Van Dyk}, S.~D. and {Lyman}, J.~D.},
        title = "{The impostor revealed: SN 2016jbu was a terminal explosion}",
      journal = {\aap},
         year = 2022,
        month = aug,
       volume = {664},
          eid = {L18},
        pages = {L18},
          doi = {10.1051/0004-6361/202244262},
archivePrefix = {arXiv},
       eprint = {2206.06365},
 primaryClass = {astro-ph.HE},
       adsurl = {https://ui.adsabs.harvard.edu/abs/2022A&A...664L..18B}
}

@ARTICLE{J22,
       author = {{Jencson}, Jacob E. and {Sand}, David J. and {Andrews}, Jennifer E. and {Smith}, Nathan and {Strader}, Jay and {Aghakhanloo}, Mojgan and {Pearson}, Jeniveve and {Valenti}, Stefano},
        title = "{Hubble Space Telescope Imaging Reveals That SN 2015bh Is Much Fainter than Its Progenitor}",
      journal = {\apjl},
         year = 2022,
        month = aug,
       volume = {935},
       number = {2},
          eid = {L33},
        pages = {L33},
          doi = {10.3847/2041-8213/ac867c},
archivePrefix = {arXiv},
       eprint = {2206.02816},
 primaryClass = {astro-ph.SR},
       adsurl = {https://ui.adsabs.harvard.edu/abs/2022ApJ...935L..33J}
}

@ARTICLE{M13,
       author = {{Mauerhan}, Jon C. and {Smith}, Nathan and {Filippenko}, Alexei V. and {Blanchard}, Kyle B. and {Blanchard}, Peter K. and {Casper}, Chadwick F.~E. and {Cenko}, S. Bradley and {Clubb}, Kelsey I. and {Cohen}, Daniel P. and {Fuller}, Kiera L. and {Li}, Gary Z. and {Silverman}, Jeffrey M.},
        title = "{The unprecedented 2012 outburst of SN 2009ip: a luminous blue variable star becomes a true supernova}",
      journal = {\mnras},
         year = 2013,
        month = apr,
       volume = {430},
       number = {3},
        pages = {1801-1810},
          doi = {10.1093/mnras/stt009},
archivePrefix = {arXiv},
       eprint = {1209.6320},
 primaryClass = {astro-ph.SR},
       adsurl = {https://ui.adsabs.harvard.edu/abs/2013MNRAS.430.1801M}
}

@ARTICLE{S11,
       author = {{Smith}, Nathan and {Li}, Weidong and {Silverman}, Jeffrey M. and {Ganeshalingam}, Mohan and {Filippenko}, Alexei V.},
        title = "{Luminous blue variable eruptions and related transients: diversity of progenitors and outburst properties}",
      journal = {\mnras},
         year = 2011,
        month = jul,
       volume = {415},
       number = {1},
        pages = {773-810},
          doi = {10.1111/j.1365-2966.2011.18763.x},
archivePrefix = {arXiv},
       eprint = {1010.3718},
 primaryClass = {astro-ph.SR},
       adsurl = {https://ui.adsabs.harvard.edu/abs/2011MNRAS.415..773S}
}

@ARTICLE{S14,
       author = {{Smith}, Nathan},
        title = "{Mass Loss: Its Effect on the Evolution and Fate of High-Mass Stars}",
      journal = {\araa},
         year = 2014,
        month = aug,
       volume = {52},
        pages = {487-528},
          doi = {10.1146/annurev-astro-081913-040025},
archivePrefix = {arXiv},
       eprint = {1402.1237},
 primaryClass = {astro-ph.SR},
       adsurl = {https://ui.adsabs.harvard.edu/abs/2014ARA&A..52..487S}
}

@ARTICLE{S10,
       author = {{Smith}, Nathan and {Miller}, Adam and {Li}, Weidong and {Filippenko}, Alexei V. and {Silverman}, Jeffrey M. and {Howard}, Andrew W. and {Nugent}, Peter and {Marcy}, Geoffrey W. and {Bloom}, Joshua S. and {Ghez}, Andrea M. and {Lu}, Jessica and {Yelda}, Sylvana and {Bernstein}, Rebecca A. and {Colucci}, Janet E.},
        title = "{Discovery of Precursor Luminous Blue Variable Outbursts in Two Recent Optical Transients: The Fitfully Variable Missing Links UGC 2773-OT and SN 2009ip}",
      journal = {\aj},
         year = 2010,
        month = apr,
       volume = {139},
       number = {4},
        pages = {1451-1467},
          doi = {10.1088/0004-6256/139/4/1451},
archivePrefix = {arXiv},
       eprint = {0909.4792},
 primaryClass = {astro-ph.SR},
       adsurl = {https://ui.adsabs.harvard.edu/abs/2010AJ....139.1451S}
}

@ARTICLE{S18,
       author = {{Smith}, Nathan and {Andrews}, Jennifer E. and {Rest}, Armin and {Bianco}, Federica B. and {Prieto}, Jose L. and {Matheson}, Tom and {James}, David J. and {Smith}, R. Chris and {Strampelli}, Giovanni Maria and {Zenteno}, A.},
        title = "{Light echoes from the plateau in Eta Carinae's Great Eruption reveal a two-stage shock-powered event}",
      journal = {\mnras},
         year = 2018,
        month = aug,
       volume = {480},
       number = {2},
        pages = {1466-1498},
          doi = {10.1093/mnras/sty1500},
archivePrefix = {arXiv},
       eprint = {1808.00992},
 primaryClass = {astro-ph.SR},
       adsurl = {https://ui.adsabs.harvard.edu/abs/2018MNRAS.480.1466S}
}

@ARTICLE{S19,
       author = {{Smith}, Nathan},
        title = "{The isolation of luminous blue variables resembles aging B-type supergiants, not the most massive unevolved stars}",
      journal = {\mnras},
         year = 2019,
        month = nov,
       volume = {489},
       number = {3},
        pages = {4378-4388},
          doi = {10.1093/mnras/stz2277},
archivePrefix = {arXiv},
       eprint = {1908.06104},
 primaryClass = {astro-ph.SR},
       adsurl = {https://ui.adsabs.harvard.edu/abs/2019MNRAS.489.4378S}
}

@ARTICLE{S22,
       author = {{Smith}, Nathan and {Andrews}, Jennifer E. and {Filippenko}, Alexei V. and {Fox}, Ori D. and {Mauerhan}, Jon C. and {Van Dyk}, Schuyler D.},
        title = "{SN 2009ip after a decade: the luminous blue variable progenitor is now gone}",
      journal = {\mnras},
         year = 2022,
        month = sep,
       volume = {515},
       number = {1},
        pages = {71-81},
          doi = {10.1093/mnras/stac1669},
archivePrefix = {arXiv},
       eprint = {2205.02896},
 primaryClass = {astro-ph.HE},
       adsurl = {https://ui.adsabs.harvard.edu/abs/2022MNRAS.515...71S}
}

@ARTICLE{GL09,
       author = {{Gal-Yam}, A. and {Leonard}, D.~C.},
        title = "{A massive hypergiant star as the progenitor of the supernova SN 2005gl}",
      journal = {\nat},
         year = 2009,
        month = apr,
       volume = {458},
       number = {7240},
        pages = {865-867},
          doi = {10.1038/nature07934},
       adsurl = {https://ui.adsabs.harvard.edu/abs/2009Natur.458..865G}
}

@ARTICLE{H94,
       author = {{Humphreys}, Roberta M. and {Davidson}, Kris},
        title = "{The Luminous Blue Variables: Astrophysical Geysers}",
      journal = {\pasp},
         year = 1994,
        month = oct,
       volume = {106},
        pages = {1025},
          doi = {10.1086/133478},
       adsurl = {https://ui.adsabs.harvard.edu/abs/1994PASP..106.1025H}
}

@ARTICLE{L98,
       author = {{Lamers}, H.~J.~G.~L.~M. and {Bastiaanse}, M.~V. and {Aerts}, C. and {Spoon}, H.~W.~W.},
        title = "{Periods, period changes and the nature of the microvariations of Luminous Blue Variables}",
      journal = {\aap},
         year = 1998,
        month = jul,
       volume = {335},
        pages = {605-621},
       adsurl = {https://ui.adsabs.harvard.edu/abs/1998A&A...335..605L}
}

@ARTICLE{H03,
       author = {{Heydari-Malayeri}, M. and {Meynadier}, F. and {Walborn}, N.~R.},
        title = "{Tight LMC massive star clusters R 127 and R 128}",
      journal = {\aap},
         year = 2003,
        month = mar,
       volume = {400},
        pages = {923-937},
          doi = {10.1051/0004-6361:20030066},
archivePrefix = {arXiv},
       eprint = {astro-ph/0301298},
 primaryClass = {astro-ph},
       adsurl = {https://ui.adsabs.harvard.edu/abs/2003A&A...400..923H}
}
\DeclareRobustCommand{\VAN}[3]{#3}
\bibliographystyle{aasjournal}



\end{document}